\documentclass[useAMS,usenatbib]{mn2e}
\usepackage{graphicx}
\usepackage{amssymb,amsmath}
\usepackage{longtable,lscape}
\usepackage{url}

\def\kmsMpc{\ifmmode {\rm\,km\,s^{-1}\,Mpc^{-1}}\else ${\rm\,km\,s^{-1}\,Mpc^{-1}}$\fi}
\def\lya{Ly$\alpha$}
\def\cf{{cf.,}}

\newcommand{\lRL}{\hbox{log~$R_{\rm L}$}}

\newcommand{\lsoft}{\hbox{$L_{\rm 2keV}$}}

\newcommand{\lIR}{\hbox{$L_{3.6 \rm \mu m}$}}
\newcommand{\llsoft}{\hbox{log~$L_{\rm 2keV}$}}
\newcommand{\llUV}{\hbox{log~$L_{2500}$}}

\newcommand{\kms}{\hbox{km~s$^{-1}$}}

\newcommand{\flux}{\hbox{erg~cm$^{-2}$~s$^{-1}$}}
\newcommand{\lumin}{\hbox{erg~s$^{-1}$}}

\newcommand{\aox}{\hbox{$\alpha_{\rm ox}$}}

\newcommand{\XR}{\hbox{X-ray}}

\newcommand{\PL}{\hbox{power-law}}

\newcommand{\RAB}{$N_{\rm AB}/N_{\rm TAR}$}

\newcommand{\chandra}{\emph{Chandra}}
\newcommand{\vlt}{VLT}
\newcommand{\vimos}{VIMOS}
\newcommand{\mos}{MOS}
\newcommand{\keck}{Keck}
\newcommand{\deimos}{DEIMOS}
\newcommand{\lris}{LRIS}

\newcommand{\spitzer}{{\emph{Spitzer}}}

\newcommand{\SB}{\hbox{0.5--2~keV}}
\newcommand{\HB}{\hbox{2--8~keV}}
\newcommand{\FB}{\hbox{0.5--8~keV}}

\newcommand\ion[2]{#1$\;${\sc{#2}}}

\def\aap{A\&A}
\def\apj{ApJ}
\def\apjl{ApJL}
\def\apjs{ApJS}
\def\aj{AJ}
\def\mnras{MNRAS}
\def\araa{ARA\&A}

\def\nat{Nature}
\def\pasp{PASP}

\title[{\bf SPECTROSCOPIC SURVEY OF THE SSA22 FIELD}]
  {An Extragalactic Spectroscopic Survey of the SSA22 Field}
\author[{\bf SAEZ ET AL.}]
  {C.~Saez,$^{1}$
  B.~D.~Lehmer,$^{2,3}$  F.~E.~Bauer,$^{4,5,6}$ D.~Stern,$^7$
 A.~Gonzales,$^8$ I.~Rreza,$^9$
  \newauthor 
  D.~M.~Alexander,$^{10}$
  Y.~Matsuda,$^{11,12}$
  J.~E.~Geach,$^{13}$
  F.~A. Harrison$^9$  and 
  \newauthor
  T. Hayashino$^{14}$ \\
  $^1$Department of Astronomy, University of Maryland, College Park, MD 20742-2421, USA\\
  $^2$The Johns Hopkins University, Homewood Campus, Baltimore, MD 21218, USA\\
  $^3$NASA Goddard Space Flight Center, Code 662, Greenbelt, MD 20771, USA \\
  $^4$Instituto de Astrof\'{\i}sica, Facultad de F\'{i}sica, Pontificia Universidad Cat\'{o}lica de Chile, 306, Santiago 22, Chile\\
  $^5$Millennium Institute of Astrophysics, Santiago, Chile\\
  $^6$Space Science Institute, 4750 Walnut Street, Suite 205, Boulder, Colorado 80301\\
  $^7$Jet Propulsion Laboratory, California Institute of Technology, Pasadena, CA 91109\\
  $^8$Department of Earth, Atmospheric and Planetary Sciences, Massachusetts Institute of Technology, Cambridge, MA 02139\\
  $^9$Cahill Center for Astrophysics, California Institute of Technology, Pasadena, CA 91125\\
  $^{10}$Department of Physics, Durham University, Durham DH1 3LE, UK\\
  $^{11}$National Astronomical Observatory of Japan, 2-21-1 Osawa, Mitaka, Tokyo 181-8588, Japan\\
  $^{12}$The Graduate University for Advanced Studies (SOKENDAI), 2-21-1 Osawa,
Mitaka, Tokyo 181-0015, Japan \\
  $^{13}$Centre for Astrophysics Research, Science \& Technology Research Institute, University of Hertfordshire, Hatfield, AL10 9AB, UK\\
  $^{14}$Astronomical Institute, Tohoku University, Aramaki, \hbox{Aoba-ku}, Sendai, Miyagi, 980-8578, Japan
  }
  
  \date{}

\pagerange{\pageref{firstpage}--\pageref{lastpage}} \pubyear{2002}

\def\LaTeX{L\kern-.36em\raise.3ex\hbox{a}\kern-.15em
    T\kern-.1667em\lower.7ex\hbox{E}\kern-.125emX}

\begin{document}

\label{firstpage}

\maketitle

\begin{abstract}

We present VLT VIMOS, Keck DEIMOS and Keck LRIS multi-object spectra of 367 sources in the field of the $z \approx 3.09$ protocluster SSA22. Sources are spectroscopically classified via template matching, allowing new identifications for 206 extragalactic sources, including  36 $z>2$ Lyman-break galaxies (LBGs) and Lyman-$\alpha$ emitters (LAEs), 8 protocluster members, and 94  \XR\ sources from the $\sim400$~ks \chandra\   deep survey of SSA22. Additionally, in the area covered by our study, we have increased by  $\approx 4$, 13, and 6 times  the number of reliable redshifts of sources  at $1.0<z<2.0$, at $z>3.4$, and with  \XR\ emission, respectively. We compare our results with past spectroscopic surveys of SSA22 to investigate the completeness of the LBGs and the \XR\ properties of the new spectroscopically-classified sources in the SSA22 field. 


\end{abstract}


\section{INTRODUCTION}\label{S:intr}

The SSA22 field hosts one of the most distant and well studied protoclusters
currently known.  Originally discovered by \cite{1998ApJ...492..428S}, the
SSA22 protocluster lies at $z = 3.09$ and contains several powerful active
galactic nuclei \citep[AGNs;][]{2009ApJ...691..687L,2009MNRAS.400..299L}, numerous Lyman-break galaxies
\citep[LBGs;][]{1998ApJ...492..428S,2003ApJ...592..728S}  and Lyman-$\alpha$ emitters \citep[LAEs;][]{2004AJ....128.2073H,2005ApJ...634L.125M}, and multiple spatially extended
Lyman-$\alpha$ blobs \citep{2000ApJ...532..170S,2004AJ....128..569M, 2011MNRAS.410L..13M}, the
most extreme of which span regions $> 100~{\rm h}^{-1}$kpc in extent.  The
protocluster has been mapped using the LAE population, which has revealed a
belt-like structure across a 60$\times$20~Mpc$^2$ (comoving size) region \citep{2004AJ....128.2073H, 2012ApJ...751...29Y}.  
At its core, the protocluster
reaches a factor of $\approx 6$ times overdensity of LBGs and LAEs compared to
the field \citep{2000ApJ...532..170S}.  Cosmological models predict that the
protocluster would have collapsed into a $z = 0$ structure resembling a rich
local cluster (e.g., Coma) with a total mass $>10^{15}M_\odot$ \citep[see, e.g.,][]{1998Natur.392..359G,1998ApJ...492..428S}.

Since the discovery of the protocluster, the SSA22 field has been extensively studied at several
wavelengths, providing new information about how galaxies grow in high density
environments at $z \approx 3$ \citep[e.g.,][]{2001ApJ...548L..17C,2004ApJ...606...85C,2005MNRAS.363.1398G,2009ApJ...692.1561W,2011ApJ...736...18N,2009Natur.459...61T,2010ApJ...724.1270T,2013MNRAS.430.2768T,2012ApJ...750..116U,2013ApJ...778..170K}.  To address how the supermassive
black hole (SMBH) populations in the protocluster environment are growing, a
deep $\approx400$~ks \chandra\ survey was conducted over the SSA22 field: the
{\it Chandra} Deep Protocluster Survey \citep{2009MNRAS.400..299L}.\footnote{See
\url{http://astro.dur.ac.uk/~dma/SSA22/} for details on the {\it Chandra} Deep
Protocluster Survey.}  The {\it Chandra} exposure over SSA22 yielded multiple
detections for LBGs and LAEs associated with the $z = 3.1$ protocluster.  Early
results indicated that the fraction of protocluster galaxies hosting AGN was
elevated by a factor of $6.1^{+10.3}_{-3.6}$ compared to the field population
at $z \approx 3$, suggesting the presence of an enhanced AGN duty cycle and/or
more massive SMBHs in the $z \approx 3$ protocluster environment \citep{2009ApJ...691..687L}. Additional AGN studies of clusters and protoclusters have
confirmed a steady rise in the AGN fraction for galaxies in the highest density
environments with increasing redshift \citep[e.g.,][]{2009ApJ...701...66M,2013ApJ...768....1M,2010MNRAS.407..846D,2013ApJ...765...87L}, implying that the key growth
phases of the most massive SMBHs occured in high-redshift protoclusters like that found in SSA22.


Prior to the current work, the \chandra\ SSA22 point source catalog from
\citet{2009MNRAS.400..299L} contained robust redshift matches for only $\approx
10\%$ of the \XR\ sources, limiting initial constraints on the AGN activity in
the protocluster to a small subset of sources in the field.
Additionally, there is only one published LBG spectroscopic catalog in the SSA22
field \citep{2003ApJ...592..728S}. This survey only covered the area around
the center of the protocluster and yielded spectroscopic measurements for
$\approx50\%$ of the known LBGs.  Therefore, in order to better characterize
the AGN activity in the SSA22 protocluster it is imperative to unambiguously identify protocluster members by obtaining 
spectroscopic measurements of both the galaxies (LBGs and LAEs) and \XR\
sources in the field.

In this paper, we present a survey of extragalactic sources in the SSA22 field
using the multi-object spectrographs at the Very Large Telescope (VLT) and
Keck. We map the SSA22 field with the dual goals of identifying new
protocluster members and complementing previous studies performed in this
field.  We target new LBGs in the redshift range of the SSA22 protocluster, as
well as \XR\ sources detected in the \chandra\ Deep Protocluster Survey.  The
structure of the paper is as follows: in $\S$2 we outline our observations and
the data reduction strategies; in $\S$3 we detail our spectral template
matching approach used to identify  the sources; in \S4 we describe the main
results; and in  \S5 we summarize our work.  Throughout this work, unless
stated otherwise, the errors listed are at the 1$\sigma$ level, we use AB magnitudes, CGS units, and adopt the concordance cosmology,
$\Omega_M = 0.3$, $\Omega_\Lambda = 0.7$ and $H_0 = 70\, \kmsMpc$.

\begin{figure}
  \includegraphics[width=8.5cm]{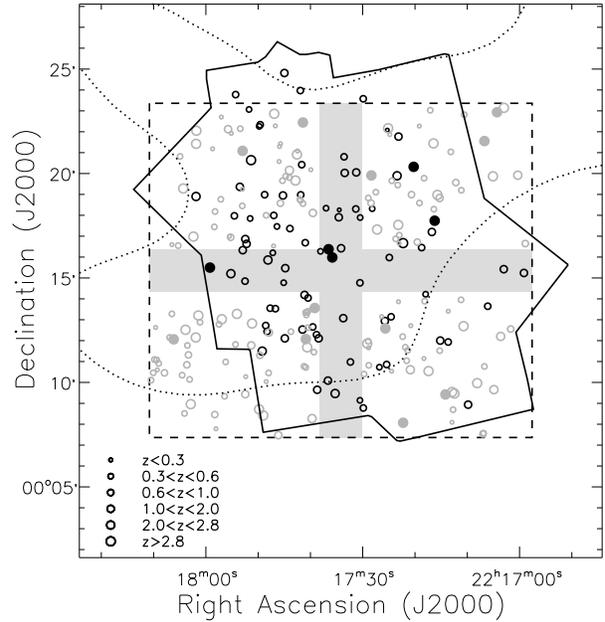}
        \caption{Positions of extragalactic sources in our sample. The black and gray circles are sources observed with \keck\ (\deimos\ and \lris), and \vlt\ \vimos, respectively.  The circles have been plotted with sizes that increase with redshift. The filled circles correspond to sources that potentially belong to the SSA22 protocluster ($3.06\leq z \leq 3.12$). The dashed square indicates the area covered by the six \vimos\  slit masks; the gray cross indicates the chips gaps. The polygon correspond to ensemble the area covered by the \keck\ slit masks. The dotted line is an isodensity contour marking where  most of the optically selected LAEs  at $z \sim 3$ reside \citep[from][]{2004AJ....128.2073H}.}
   \label{fig:ours}
     \end{figure}

\section{OBSERVATIONS AND DATA REDUCTION} \label{S:data}

\begin{table*}
\begin{minipage}{175mm}
\begin{center}
\caption{Log of Keck and VLT spectroscopic observations.. \label{tab:mask}}
\begin{tabular}{lcccccc}
\hline\hline
{\sc mask id} &
{\sc instrument} &
{\sc obs. date} &
{\sc RA} &
{\sc DEC} &
{\sc P.A. }  & 
{{\sc n}$_{\rm new}$/{\sc n}$_{\rm rep}$}  \\
\hline
SSA22a (Keck1) & DEIMOS & 2009 Sep 16 & 22:17:42.74 & 00:16:44.7 & $-$170.9 & 27/0  \\
SSA22b (Keck2)  & DEIMOS & 2009 Sep 16 & 22:17:40.27 & 00:15:20.2 & $-$133.5 & 27/5 \\
SSA22c (Keck3)  & DEIMOS  & 2009 Sep 17 & 22:17:17.05 & 00:16:32.8 & $-$166.4 & 19/1  \\
SSA22d (Keck4)  & DEIMOS  & 2009 Sep 17 & 22:17:25.64 & 00:18:41.5 & $-$128.0 & 16/1 \\
SSA22e (Keck5) & LRIS & 2009 Dec 15 & 22:17:33.53 & 00:18:36.8 & +106.0 & 6/1 \\
SSA22f (Keck6)  & LRIS & 2009 Dec 16 & 22:17:44.20 & 00:14:56.0 & +99.1 & 7/2 \\
2096 (VLT1) & VIMOS & 2011 Jun 09 & 22:17:34.60 & 00:15:20.0  & +90.0 & 98/0 \\
2318 (VLT2) & VIMOS & 2011 Sep 20 & 22:17:34.60 & 00:15:20.0  & +90.0 & 100/0  \\  
2349 (VLT3) & VIMOS & 2011 Sep 23 &  22:17:34.60 & 00:15:20.0  & +90.0 & 53/6   \\
2523 (VLT4) & VIMOS & 2012 Jul 22 & 22:17:34.60 & 00:15:20.0  & +90.0  & 11/64   \\
2526 (VLT5) & VIMOS & 2012 Aug 15 & 22:17:34.60 & 00:15:20.0  & +90.0 & 10/54  \\ 
2531 (VLT6) & VIMOS & 2012 Aug 22 & 22:17:34.60 & 00:15:20.0  & +90.0 & 5/27   \\ \hline
\end{tabular}
\end{center}
%
Characteristics of the Keck and VLT slit masks observed in the SSA22 field, including the instruments, observation dates (UT), and mask center coordinates (J2000.0) and position angles. ÊThe number of new spectroscopic targets and those that have previously determined redshifts have been indicated as $N_{\rm new}$ and $N_{\rm rep}$, respectively.
\end{minipage}
\end{table*}

Our observations were conducted using multiple slit mask exposures of the Visible Multi-Object Spectrograph \citep[\vimos;][]{2003SPIE.4841.1670L} on the \vlt, and the Deep Imaging Multi-Object Spectrograph \citep[\deimos;][]{2003SPIE.4841.1657F} and Low Resolution Imaging Spectrometer \citep[\lris;][]{1995PASP..107..375O} on \keck. These observations cover an important fraction of the areal footprint of the SSA22 \chandra\ field (see Figure~\ref{fig:ours}). In some observations (Keck LRIS, VLT VIMOS), we obtained flux calibrated spectra through the use of response curves obtained from standard stars. In these cases, the spectra have been corrected by atmospheric extinction. Although Keck DEIMOS and VLT VIMOS lack an atmospheric dispersion compensator (ADC), we do not attempt atmospheric refraction corrections. This is because observations performed by these instruments are executed at small zenith angles with N-S oriented slits, and consequently, with negligible dispersion effects \cite[see e.g.,][]{1982PASP...94..715F, 2005A&A...443..703S,  2014A&A...566A...2S}. When flux calibrated spectra are provided, we do not attempt to correct the spectra for slit losses \cite[see e.g.,][]{2011A&A...525A.143C}. Therefore, in cases of flux calibration, the spectra are accurate to the 10$-$20\% level.

\begin{table*}
\begin{minipage}{175mm}
\begin{center}
\caption{Spectroscopically surveyed extragalactic sources. \label{tab:spsr}}
\begin{tabular}{cccccccccccccccc}
\hline\hline
{\sc object id} &
{\sc RA$^{\rm a}$} &
{\sc DEC$^{\rm a}$} &
{$z$}  &
{\sc sdss $g^{\rm b}$} &
{\sc subaru $R^{\rm b}$} &
{\sc ukidss $K^{\rm b}$} &
{\sc s. type$^{\rm c}$}  &
{\sc masks$^{\rm d}$}  &
{\sc quality} &
{\sc class$^{\rm e}$} \\
\hline
\multicolumn{11}{c}{\small  VLT observations} \\
 \\
\hline
   10031 & 334.27914 &  0.12557 &  0.6292 &  23.0 &  21.3 &  18.6 &    EllG &   1,4 &  A &           bri \\
   10032 & 334.30383 &  0.13486 &  0.1000 &  20.4 &  21.1 &  20.2 &    EllG &     1 &  B &           bri \\
   10034 & 334.49194 &  0.12968 &  0.1718 &  21.1 &  21.3 &  20.0 &     SFG &   1,4 &  A &           bri \\
   10051 & 334.30301 &  0.14493 &  0.3839 &  22.7 &  21.1 &  18.8 &    EllG &   3,5 &  A &           bri \\
   10077 & 334.45193 &  0.16735 &  0.1102 &  21.4 &  21.9 &  21.4 &     SpG &   1,5 &  A &           bri \\
   10092 & 334.51294 &  0.16906 &  0.3034 &  21.8 &  21.5 &  19.5 &     SpG &   1,4 &  A &           bri \\
   10110 & 334.42041 &  0.17862 &  0.3986 &  22.7 &  22.0 &  20.4 &     SpG &   1,6 &  A &           bri \\
   10113 & 334.53757 &  0.18189 &  0.4213 &  23.0 &  21.8 &  20.3 &     SpG &   1,6 &  A &           bri \\
   \hline
\\
\multicolumn{11}{c}{\small  Keck observations} \\
 \\
 \hline
     006 & 334.24649 &  0.25400 &  1.1261 &  22.1 &  21.9 &  20.1 &     AGN &     4 &  A &        XR/bri \\
     010 & 334.26257 &  0.25703 &  1.0252 &   ... &  24.9 &  20.2 &     AGN &     4 &  B &            XR \\
     016 & 334.27536 &  0.22747 &  0.9070 &   ... &  24.7 &  21.7 &     SpG &     4 &  A &            XR \\
     021 & 334.29099 &  0.14894 &  1.1123 &   ... &  25.3 &  21.5 &     SpG &     3 &  B &            XR \\
    035p & 334.30688 &  0.19892 &  0.7532 &  23.4 &  23.3 &  20.4 &     SpG &     3 &  A &            XR \\
     038 & 334.31320 &  0.20006 &  1.7338 &  22.5 &  22.5 &  20.9 &     AGN &     3 &  A &        XR/bri \\
     039 & 334.31378 &  0.31386 &  1.3997 &  22.1 &  21.7 &  20.5 &     AGN &     3 &  A &        XR/bri \\
    043p & 334.31766 &  0.29569 &  3.0980 &   ... &  24.5 &   ... &     AGN &     3 &  A & XR/$z$$\sim$3 \\
     \hline
\end{tabular}
\end{center}

$^{\rm a}$Optical positions in J2000.0 equatorial coordinates. \\
$^{\rm b}$The optical magnitudes presented in this Table come from the SDSS survey \citep[e.g.,][]{2008ApJS..175..297A}, SSA22 photometric survey of \cite{2004AJ....128.2073H} (Subaru magnitudes) and the UKIDSS survey \citep[e.g.,][]{2007MNRAS.379.1599L}. The UKIDSS magnitudes have been transformed from Vega to AB magnitudes using K(AB)=K(Vega)+1.9 \citep[from][]{2006MNRAS.367..454H}. \\
$^{\rm c}$Spectral types are based on cross-correlation template that has minimum $\chi^2$ (see details in \S\ref{S:span}). EllG $\equiv$ elliptical galaxy;  SpG $\equiv$ spiral Galaxy; SFG $\equiv$ starburst galaxy; LBG $\equiv$ Lyman-break galaxy; AGN $\equiv$ active galactic nuclei.\\
$^{\rm d}$ Masks where source presents A or B quality spectra.\\
$^{\rm e}$ Target selection criteria classes. bri $\equiv$ bright source ($R<22.5$); XR $\equiv$ X-ray source from the \cite{2009MNRAS.400..299L} catalog; $z$$\sim$3 $\equiv$ $z\sim3$ LBG; $z$$\sim$4 $\equiv$ $z\sim4$ LBG; Ste03 $\equiv$ LBG from \cite{2003ApJ...592..728S}; LAE $\equiv$ LAE from \cite{2004AJ....128.2073H}.  \\
Table~\ref{tab:spsr} is presented in its entirety in the electronic version (also at \url{ftp://cdsarc.u-strasbg.fr/pub/cats/J/MNRAS/450/2615}); an abbreviated version of the table is shown here for guidance as to its form and content.\\
\end{minipage}
\end{table*}

\subsection{VLT observations}
The \vimos\ observations were centered at the coordinates ${\rm 22^h17^m34.20^s}$, $0^\circ15\arcmin21.8\arcsec$ and span an area of 18\arcmin$\times$16\arcmin\ (see Figure~\ref{fig:ours}). These observations were performed in \mos\ (multi-object spectroscopy) mode using the \hbox{\sc hr blue} grism. The  wavelength coverage of 3700--6700~\AA\ covers AGN emission lines (e.g., Ly$\alpha$, \ion{C}{iv} and \ion{C}{iii}) at $1.5\lesssim z \lesssim 4$. The observations consisted of six slit masks, each with a total integration time of $\sim4$~hrs for masks used in 2011 (masks 1--3;  see Table~\ref{tab:mask} for details) and $\sim6$~hrs for masks used in 2012 (masks 4--6; Table~\ref{tab:mask}).  The exposures times were selected in order to detect faint optical counterparts to limiting magnitudes of $R\approx25$ for extragalactic sources with strong emission features and $R\approx24$  for sources with moderate emission/absorption features.
The spectra on each mask are reduced using the standard ESO \vimos\ pipeline version 2.9.7.\footnote{\url{http://www.eso.org/sci/software/pipelines}} This pipeline consist of routines from the VIMOS Interactive Pipeline and Graphical Interface \citep[VIPGI;][] {2005PASP..117.1284S}. The conditions were photometric, with $\lesssim1\arcsec$ seeing and the resolving power of these observations is ${\mathcal R}  \equiv \lambda / \Delta \lambda \sim 2300$.
Each \vimos\ mask had $\sim 150$ slits, of which $\sim 50\%$ provide spectra of sufficient quality for our template matching (see \S \ref{S:span}). 
The spectra obtained from the \vimos\ pipeline contained some contamination of emission lines due to the spectral multiplexing and sky lines. The ranges where this contamination is present has been masked out from the spectra in order to perform cross-correlation template fits (see \S\ref{S:span}). The extracted \vimos\ spectra were flux-calibrated using standard response curves\footnote{\url{http://www.eso.org/observing/dfo/quality/VIMOS/qc/response.html}} obtained from repeated exposures of standard stars. The flux calibration has been further refined by comparing the brightness of stars observed in our survey (see Appendix~\ref{S:STAR})  with their SDSS photometry.

\subsection{Keck observations}
In 2009 September, we observed four slit masks (SSA22a-d) targeting \XR\ sources in the SSA22 field with \deimos\ on the \keck~II telescope.  The \deimos\ masks typically each contained 75 targets and were observed for 1.5--2~hr, split into 20--30~min integrations; details of the observations are presented in Table~\ref{tab:mask}. 
All observations used 1\farcs5 wide slitlets and the 600~line~mm$^{-1}$ grating ($\lambda_{\rm blaze} = 7500$ \AA; ${\mathcal R} \sim 1600$).  The conditions were photometric, with 0\farcs5 seeing.  The wavelength coverage of these observations was 4600--9700~\AA\ with a spectral resolution of $\approx4$~\AA.
In 2009 December we observed two slit masks (SSA22e and f) targeting SSA22 \XR\ sources using \lris, a dual-beam spectrograph on the \keck~I telescope.  Each mask was observed for $\sim 1$~hr, split into three exposures.  All observations used 1\farcs3 wide slitlets, the 400 line~mm$^{-1}$  blue grism ($\lambda_{\rm blaze} = 3400$ \AA; ${\mathcal R} \sim 600$), and the 400~line~mm$^{-1}$ red grating ($\lambda_{\rm blaze} = 8500$ \AA; ${\mathcal R} \sim 700$).  Slit mask SSA22e, observed the first night, used the 5600~\AA\ dichroic, while SSA22f, observed the following night, used the 6800~\AA\ dichroic.  Conditions were clear for the \lris\ observations. When LRIS spectra were observed in both the blue and the red grism, we obtained a wavelength coverage of  3500--10000~\AA\ with a spectral resolution of  $\approx10$~\AA.
The \deimos\ data were processed using a slightly modified version of the pipeline developed by the DEEP2 team at UC-Berkeley \citep{2012ascl.soft03003C, 2013ApJS..208....5N}, while the \lris\ data reduction followed standard procedures.  


Given that our targets are \XR\ sources and high-redshifts galaxies primarily, in the next sections we focus our analysis mainly on the extragalactic sources. In Appendix~\ref{S:STAR} we present information related to stellar sources observed in the field.

\section{SPECTRAL ANALYSIS} \label{S:span}

We targeted sources for spectroscopy based on three target selection criteria classes.
The first category corresponds to sources in the  \chandra\ \XR\ catalog of the SSA22 field  from \cite{2009MNRAS.400..299L}. The second category includes bright sources with optical magnitudes $R < 22.5$.\footnote{The $R$-magnitudes come from the Subaru photometric survey of  \cite{2004AJ....128.2073H}} The third category corresponds to optically selected LBG  and LAE candidates, further divided into four sub-categories: \cite{2003ApJ...592..728S} $z\sim3$ LBG candidates, new $z\sim 3$ or  $z\sim 4$  LBG candidates (see \S \ref{S:LBGs} for details), and LAE candidates from \cite{2004AJ....128.2073H}. The majority of \keck\ targets correspond to sources in the \XR\ catalog of  \cite{2009MNRAS.400..299L}.  

   \begin{figure}
   \includegraphics[width=8.5cm]{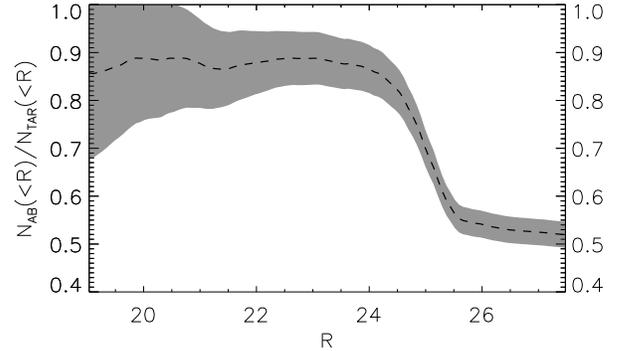}
        \centering
       \caption{Ratio of the cumulative number A or B quality spectra to the cumulative number of total targets as a function of $R$ magnitude. The shadow area are the 1-$\sigma$ errors (Poisson errors are assumed).}
     \label{fig:RTAR}
     \end{figure}

To determine redshifts and spectral types for our observed objects, we fit our spectra with several templates using the publicly available software {\sc specpro} \citep{2011PASP..123..638M}. 
The extragalactic templates used (see Appendix~\ref{S:STAR} for a description of the stellar templates) come from the VVDS \citep{2013A&A...559A..14L} for Spiral (SpG), Elliptical (EllG), and Star-Forming Galaxies (SFG); \cite{2003ApJ...588...65S} and \cite{2013MNRAS.430..425B} for LBGs in emission and absorption; \cite{1991ApJ...373..465F} for Seyfert galaxies; and \cite{2001AJ....122..549V} for QSOs. For each spectrum, we select a cross-correlation template based on the minimum $\chi^2$. In general, almost every recognizable extragalactic spectrum will be well described by one of the templates. 
 Therefore, based on the best-matched template we obtain redshifts and classify the spectra respectively as SpG, EllG, SFG, AGN (for QSO or Seyfert), or LBG. 
 
 In Table~\ref{tab:spsr}, we present basic information for our observed objects. This table summarizes the results of our source spectroscopy, including the quality of the derived spectroscopic redshift ($Q$).
Quality flag $Q={\rm A}$ signifies an unambiguous redshift determination,
typically relying on an asymmetric line profile indicating \lya\
emission \citep[\cf][]{2000ApJ...537...73S}, a confidently identified continuum
break in the spectroscopy (e.g., a Lyman-break or D4000 break), and/or
the identification of multiple emission/absorption features.  Quality
flag $Q={\rm B}$ signifies a less certain redshift determination, where
the asymmetry of the emission line was ambiguous, the continuum
break was of uncertain identification, and/or only a single emission
feature was identified.  We consider $Q={\rm B}$ spectroscopic
redshifts to be likely correct; however, additional spectroscopy would
be beneficial. Some of the sources have been observed multiple times.
In the last column of Table~\ref{tab:mask}, $N_{\rm new}$ is the number of previously unobserved sources on a given mask, and $N_{\rm rep}$ is the number of sources which were also observed on a previous mask. $N_{\rm new}$ and $N_{\rm rep}$ account for quality A and B obtained in this work, including both extragalactic sources and stars. For cases of spectral quality A or B, the redshift errors are conservatively expected to be $\Delta z \lesssim 0.0005(1+z)$, based on the spectral resolution combined with uncertainties in the wavelength calibration. These errors are verified by comparing the redshifts obtained for sources with good quality spectra in more than one mask. 
For cases where the extragalactic sources  were observed in more than one slit, we also checked the reliability of the redshift determinations obtaining cases where $\Delta z < 0.0025 (1+z)$. For $Q={\rm A}$ and $Q={\rm}B$ spectra the reliability is  98\% (62 out of 63 cases) and 97\% (28 out of 29 cases), respectively. 
In Figure~\ref{fig:RTAR} we plot the ratio of the cumulative number of A/B quality spectra to the total number of targeted slits (\RAB) as a function of magnitude in the $R$-band. This ratio is more or less constant at $\sim 90$\% with increasing $R$, up to $R\sim24$. As expected, for $R>24$ \RAB\ declines dramatically.
Table~\ref{tab:spsr} includes the target selection criteria class; many of the sources that were previously classified as bright and a few of the \XR\ sources were found to be stars (see Appendix~\ref{S:STAR} for details of stellar spectra). The ratio of A/B spectra to the number of targeted sources (\RAB) of each target class is as follows (Poisson errors are assumed): $0.89\pm0.06$ for bright, $0.52\pm 0.05$ for \XR, $0.28\pm0.04$ for $z\sim3$ LBGs, and $0.17\pm0.05$ for $z\sim4$ LBGs (see Table~\ref{tab:sura} for details).

 \begin{figure*}
\includegraphics[width=16cm]{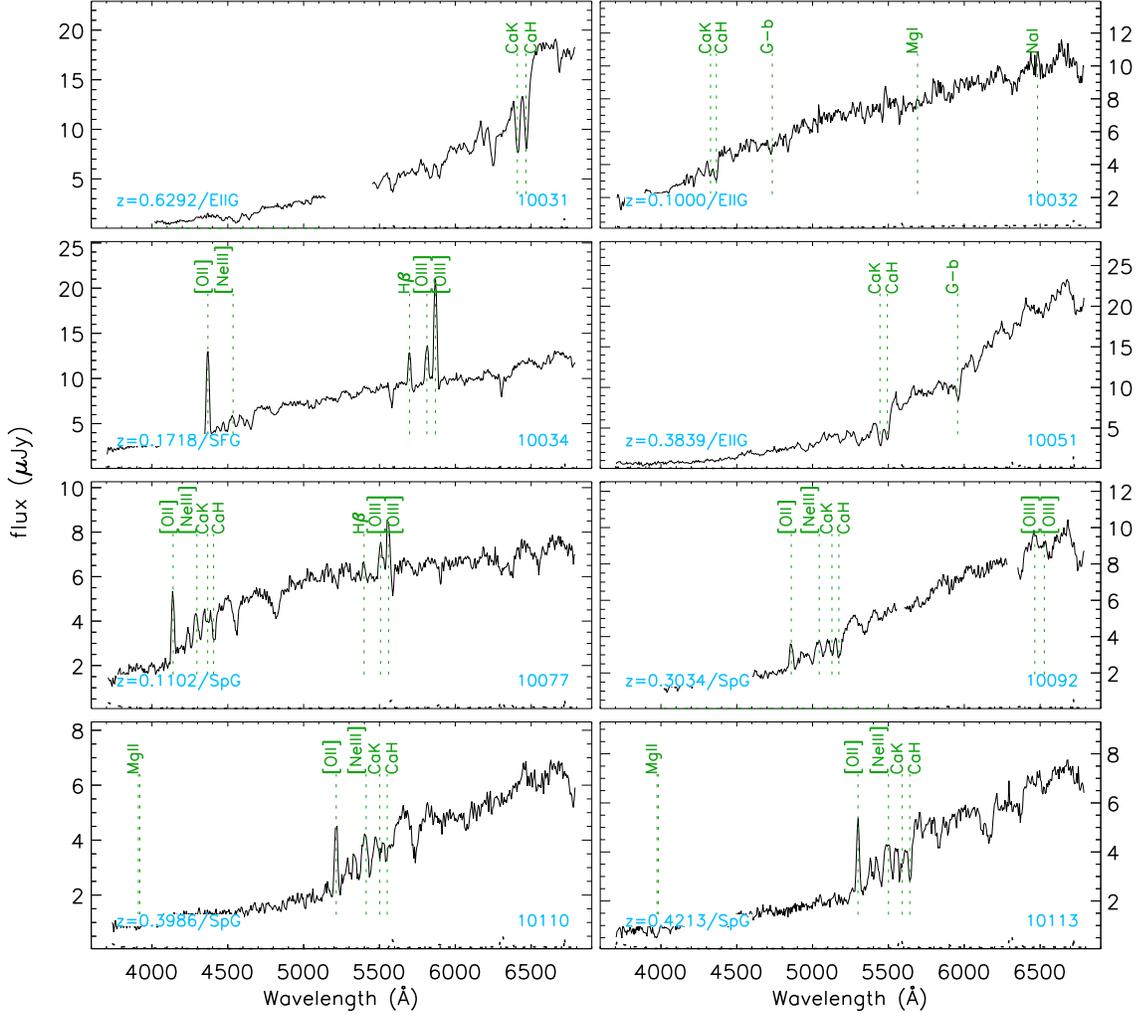}
\caption{Sample \vlt\ \vimos\ spectra. In each figure key absorption and emission features of the corresponding template used to fit the spectrum are marked for reference. For display purposes only the first eight \vimos\  spectra are shown. Please refer to the electronic version to see all the spectra.}
\label{fig:spGA}
\end{figure*}

\begin{figure*}
\includegraphics[width=16cm]{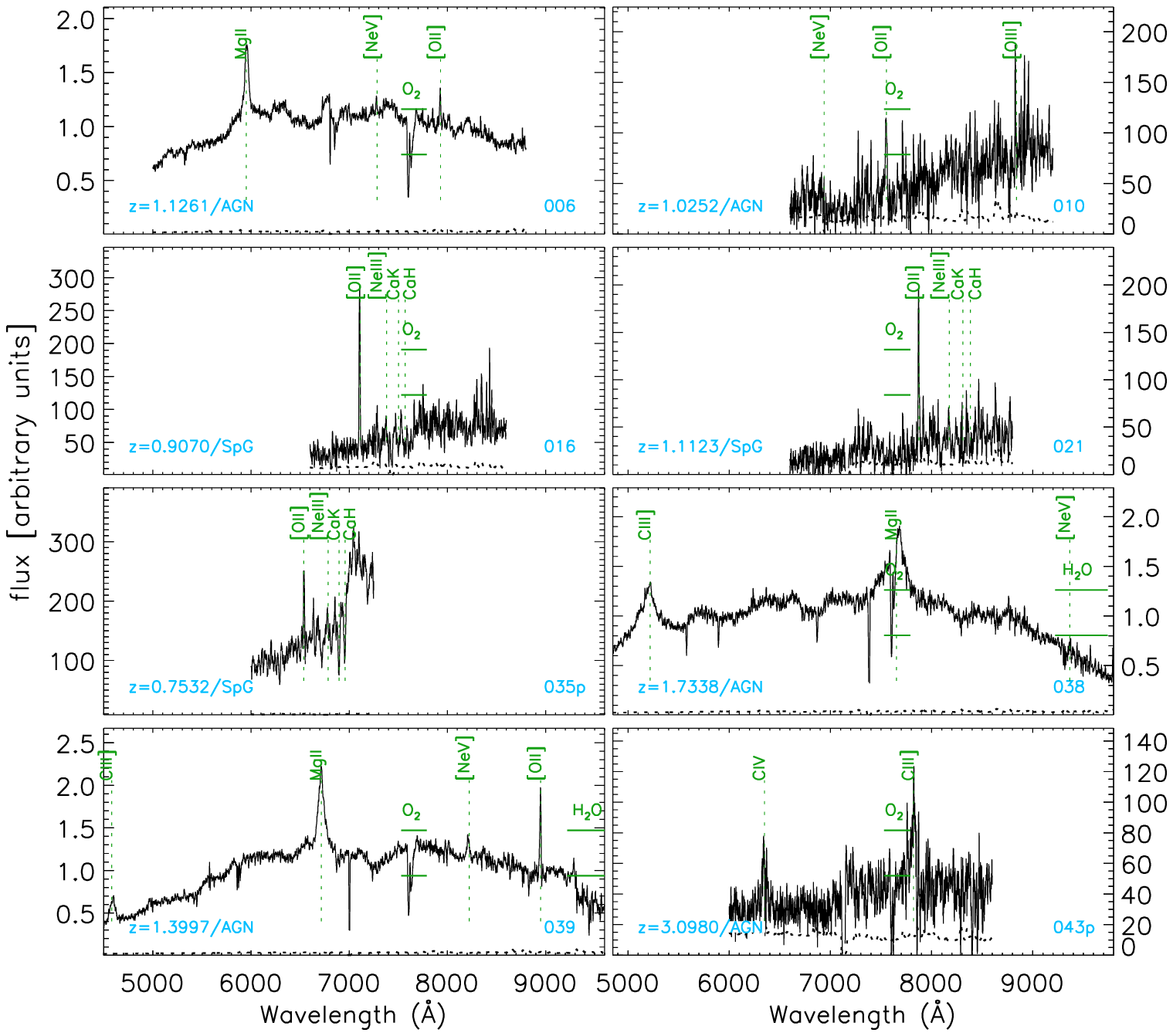}
\caption{Sample Keck \deimos\ and \lris\ spectra. In each figure key absorption and emission features of the corresponding template used to fit the spectrum are marked for reference.  For display purposes only the first eight Keck spectra are shown. Please refer to the electronic version to see all the spectra.}
\label{fig:spG2}
\end{figure*}
     
\begin{table}
\begin{center}
\caption{The ratio of A/B spectra to the number of targeted sources. \label{tab:sura}}
\begin{tabular}{lccc}
\hline\hline
class$^{\rm a}$ &  
VLT obs.  & 
Keck obs. & 
VLT \& Keck obs. \\
&  
\RAB  & 
\RAB & 
\RAB \\ 
\hline
Bright & 218/246 & 38/41 & 245/274 \\ 
\XR &  35/94 & 97/186 & 120/233 \\ 
$z\sim3$ LBG & 38/140 & 2/2 & 39/141 \\
$z\sim4$ LBG &  12/70 & 0/1 & 12/71 \\ 
All LBG &  52/254 & 6/11 & 57/262 \\ 
All classes & 277/561 & 102/200 & 367/714 \\
\hline
\end{tabular}
%
\end{center}
$^{\rm a}$ Target selection criteria class.\\
$N_{\rm AB}$ is the number of A or B quality spectra and $N_{\rm TAR}$ is the total number of targeted sources.\\

\end{table}

In Figures~\ref{fig:spGA} and \ref{fig:spG2} we plot flux/counts as a function of wavelength for the \vlt\ and \keck\  spectra of extragalactic sources respectively; identified absorption and emission lines of the best matched spectral template are indicated. 
The spectral classification for most sources was based on templates that match the source spectra for at least two noticeable absorption/emission features. The exceptions to this are the LBGs  (see \S\ref{S:LBGs} for more details). In total we have spectroscopically classified (with A/B quality) 164 extragalactic sources and 113 stars in the \vlt\ observations, and 94  extragalactic sources and 8 stars in the \keck\ observations. In section~\ref{S:PeSp}, we describe in more detail the spectra of three sources that show peculiar spectral features. Discounting overlaps  (matching radius $\leq 1\farcs3$) between the \vlt\ and \keck\ observations, we have obtained reliable spectra for a total of 247 unique extragalactic sources and 120 unique Galactic stars.  For  sources observed by both the \vlt\ and \keck\ observations,  we found consistency in the wavelength range where the spectra overlapped.


\section{RESULTS AND DISCUSSION}  \label{S:resu}

\subsection{Comparison with Published Redshifts} \label{S:surv}
We compare our redshifts with those from published spectroscopic surveys that include the SSA22 field.  These surveys include the \cite{2008A&A...486..683G}  \vimos\ \vlt\ Deep Survey \citep[VVDS;][]{2005A&A...439..845L}, the LAE FOCAS Subaru surveys by \cite{2005ApJ...634L.125M} and \cite{2012ApJ...751...29Y},  and the Keck LBG survey by \cite{2003ApJ...592..728S}. We selected sources in those surveys that were within an area of 23\arcmin$\times$23\arcmin\  centered on the \vlt\ survey center.  The selected area is large enough to cover the LBG fields SSA22a from \cite{2003ApJ...592..728S}  and the \chandra\ survey from the SSA22 field \citep{2009MNRAS.400..299L}. In the published surveys, we select sources with redshift quality similar to those of our surveys. In the case of the VVDS survey, we select only flags 3 and 4, which indicates a redshift reliability above 95\%. Additionally in the \cite{2003ApJ...592..728S} surveys, we selected only secure redshift identifications of optically selected LBG candidates with a strong emission feature (identified as a Ly$\alpha$ emission line) and/or cases of multiple absorption features. Therefore, the spectral quality of the selected sources in \cite{2003ApJ...592..728S} is very similar to $Q={\rm A}$ or $Q={\rm B}$ spectra from this work (see \S \ref{S:LBGs}). Finally, the LAE redshift identifications of \cite{2005ApJ...634L.125M} and \cite{2012ApJ...751...29Y} correspond to cases of a high S/N emission feature often complemented with multiple absorption features. Therefore, the selected LAE redshift identifications are equivalent to $Q=A$ or $Q=B$ spectra from this work as well (see \S \ref{S:LBGs}). 
Using this criterion, there are 494 source redshifts available in published catalogs, including 281 reliable redshifts from the VVDS survey, 100 sources from  \cite{2003ApJ...592..728S}, and 113 sources from the LAE surveys. Note that we do not count sources that are repeated  across the published surveys. For repeated sources (angular distance $< 1\farcs3$), we have taken the following arbitrary order of priority (from highest to lowest priority): the \cite{2003ApJ...592..728S}  LBG survey, the LAE survey of \cite{2012ApJ...751...29Y}, the LAE survey of  \cite{2005ApJ...634L.125M}, and then the VVDS survey.

\begin{figure}
   \includegraphics[width=8.8cm]{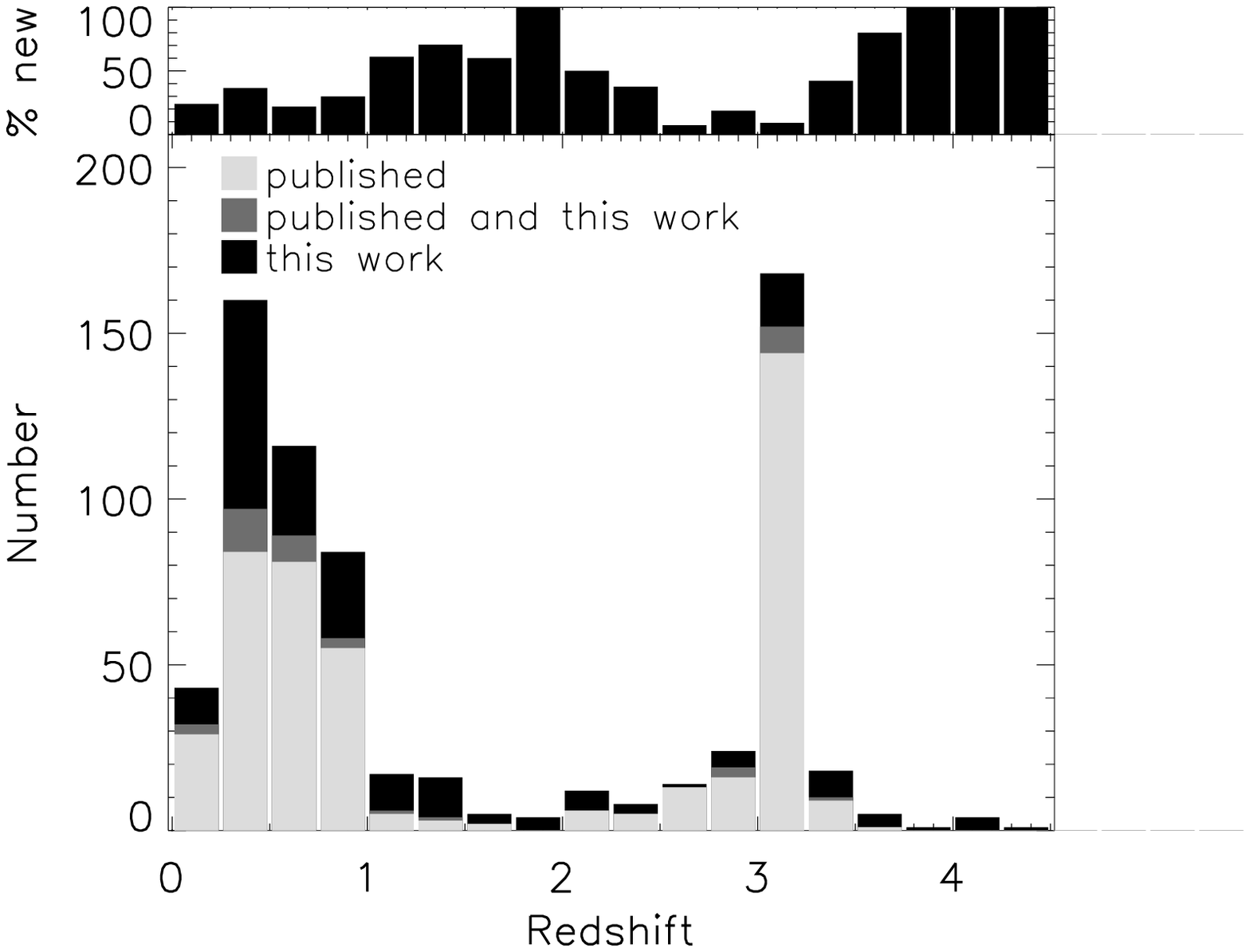}
        \centering
       \caption{Redshift distribution of sources in the SSA22 field. These sources have been selected inside a 23\arcmin$\times$23\arcmin\  square centered on the coordinates ${\rm 22^h17^m34.20^s}$, $0^\circ15\arcmin21.8\arcsec$ (J2000). The dark and light gray histograms represent sources collected by published redshift surveys with and without overlap with our surveys. The black histogram represents sources with new redshifts from our survey. The upper panel indicates the redshift distribution of the fraction of new sources coming from our surveys.}
     \label{fig:hist}
     \end{figure}

The redshift distribution of our sample is compared with those available in the literature in Figure~\ref{fig:hist}. The VVDS survey provides the majority of redshifts at $z \lesssim 1.0$, while the other surveys dominate the source redshift distribution at $z>2.0$.  By design, the LAE surveys are very concentrated on sources at $z\approx 3.0$ (see \S \ref{S:prot}). 
Our spectroscopy included 41 sources that already had redshifts from the public surveys  (matching radius $< 1\farcs3$).  In these sources we do not find major differences between our redshifts and published redshifts ($\Delta z < 0.005(1+z)$). Therefore, we spectroscopically identified 206 (247 total minus 41 previously classified) new extragalactic sources, a $\approx 42$\% increase in the number of reliable spectroscopic redshifts.  Our major contribution comes in the redshift range between $1.0 \leq z \leq 2.0$ and $z \geq 3.4$, where we have increased the number of sources with redshifts by factors of $3.5$ (from 12 to 42) and 13 (from 1 to 13), respectively. The sources at $1.0 \leq z \leq 2.0$ come mainly from the \keck\ observations, since DEIMOS and LRIS probe wavelengths up to $\approx 1\mu {\rm m}$. The sources at $z \geq 3.4$ were obtained by targeting LBGs candidates at $z\sim4$ with \vlt-\vimos\  (see \S \ref{S:LBGs}).

\subsection{Lyman-Break Galaxies\label{S:LBGs}}

\begin{figure}
   \includegraphics[width=8.8cm]{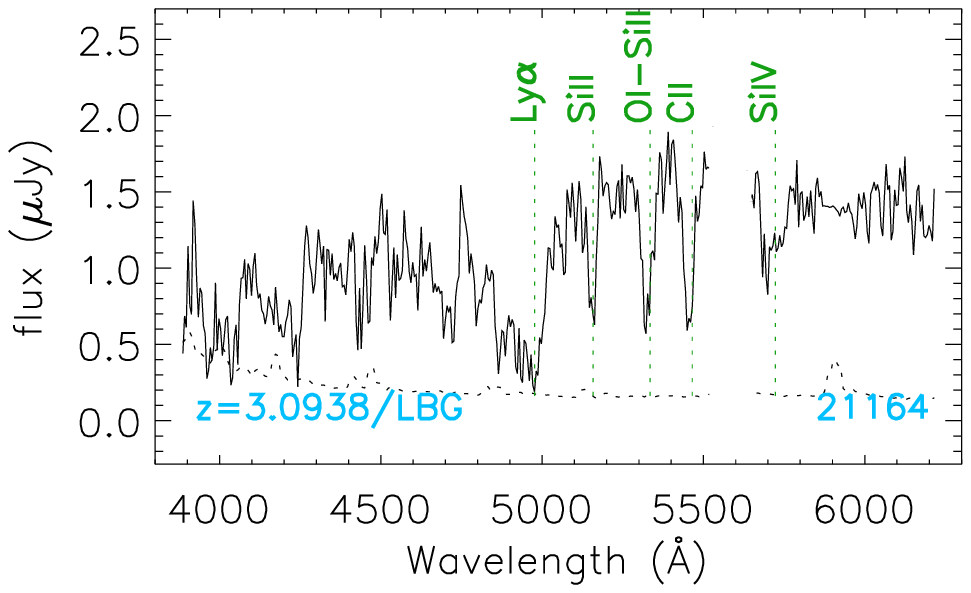}
        \centering
       \caption{LBG absorption spectrum of VLT source 21164. Some absorption lines of the LBG absorption template have been shown for reference.}     \label{fig:LBGA}
     \end{figure}

Our LBG template was used to fit $z > 2$ galaxies that have been optically classified either as LBGs or LAEs.
The majority of  LBGs  have been previously classified in four groups: $z\sim3$ LBGs, $z\sim4$ LBGs \citep[from the photometric survey of][]{2004AJ....128.2073H}, optical LBGs candidates from \citet{2003ApJ...592..728S}, and optical LAEs candidates from \cite{2004AJ....128.2073H}. 

\vspace{4pt}
The criteria to select $z\sim3$ LBGs was:

\vspace{2pt}
\begin{tabular}{l l l }
(1) $22 \leq R \leq 25.5$  & \hspace{0.0cm}  & (4) $V-R \leq 0.3$ \\
(2) $U-V-2(V-R) \geq 1.0$ & & (5) $R-z \leq 0.2$ \\
(3) $U-V \geq1.0$ & & \\
\end{tabular}

\vspace{6pt}
The criteria to select $z\sim4$ LBGs was:

\vspace{2pt}
\begin{tabular}{l l l }
(1) $U({\rm CFHTLS})>27.2$ & \hspace{0.0cm}  & (5) $B-i\geq1.9$ \\
(2)  $B-i-2.3(i-z)\geq1.9$ & & (6) $i-z\leq0.15$ \\
(3) $V-i-1.61(i-z)\geq0.5$ & & (7) $i\leq26.0$ \\
(4) $R-i\leq1.2$ & &  \\
\end{tabular}

\vspace{6pt}

Note that in the last expressions CFHTLS stands for Canada-France-Hawaii Telescope. With the exception of one source (\vimos\ source 21164; see Figure~\ref{fig:LBGA} and Table~\ref{tab:spsr}),  the LBGs show clear evidence of a single strong emission line, assumed to be Ly$\alpha$.  \vimos\ source 21164  ($R=23.7$) was the only case found with Ly$\alpha$ absorption.
This feature was accompanied by at least three absorption lines,  allowing a very reliable cross correlation match with our LBG absorption template (see Figure~\ref{fig:LBGA}). Among the absorption lines observed the spectrum of source 21164 are: \ion{Si}{ii}~$\lambda$1260, \ion{O}{i}+\ion{Si}{ii}~$\lambda$1303 and \ion{C}{ii}~$\lambda$1334.
The \vlt\ (\keck) observations showed evidence for high-redshift ($z>2$) LBG spectral signatures in 43 (2) sources. Discounting overlaps with published redshifts, we found a total of 36 new sources with LBG spectral signatures (including LBGs and LAEs). 
Since the \keck\ observations exclusively targeted  \chandra\ \XR\ sources, all LBGs from \keck\ are \XR\ emitters. 

\begin{table*}
\begin{minipage}{175mm}
\begin{center}
\caption{Distribution of $z\sim3$ and $z\sim4$ LBG candidates in two magnitude bins.\label{tab:LBGa}}
\begin{tabular}{lccccc}
\hline\hline
 &  
 & 
 \multicolumn{2}{c}{$z\sim3$} & 
 \multicolumn{2}{c}{$z\sim4$} \\
Sources & 
With Spectroscopic $z$ & 
$R<25$ & 
$R\geq25$ & 
$R<25$& 
$R\geq25$ \\
\hline
Published and not in this work$^{\rm a}$ & Yes & 37 & 17 & 2 & 1 \\ 
Published and in this work$^{\rm b}$ & Yes & 3 & 4 & 0 & 0  \\ 
This work only$^{\rm c}$ & Yes & 14 & 18 & 6 & 6  \\
Not $Q$=A, B targets with redshifts$^{\rm d}$ & Yes & 5 & 2 & 1 & 0 \\
Not $Q$=A, B  targets with no redshift$^{\rm e}$ & No & 31 & 64 & 15 & 43 \\ 
No information & No  &  80 & 181 & 58 & 363 \\ 
\hline
\end{tabular}
\end{center}
The area where the sources were selected correspond to a  23\arcmin$\times$23\arcmin\  square centered on the coordinates ${\rm 22^h17^m34.20^s}$, $0^\circ15\arcmin21.8\arcsec$ (J2000).\\
$^{\rm a}$ Sources with published redshifts and with no $Q=$~A or B redshifts from this work.\\
$^{\rm b}$ Sources with published redshifts and with $Q=$~A or B redshifts from this work.\\
$^{\rm c}$ Sources with  $Q=$~A or B redshifts from this work only.\\
$^{\rm d}$ Targets from this work that did not result in $Q$=A or B spectra and with redshifts from published surveys.\\
$^{\rm e}$ Targets from this work that did not result in $Q$=A or B spectra and with no redshifts from published surveys.
\end{minipage}
\end{table*}

\begin{figure}
   \includegraphics[width=8.8cm]{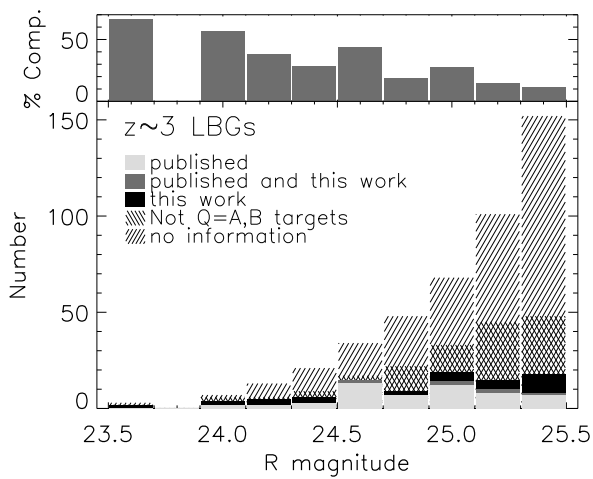}
        \centering
       \caption{Magnitude distribution of $z\sim 3$ optically selected LBG candidates in the SSA22 field. These sources have been selected inside a 23\arcmin$\times$23\arcmin\  square region centered on the coordinates ${\rm 22^h17^m34.20^s}$, $0^\circ15\arcmin21.8\arcsec$ (J2000). 
       The dark and light gray histograms represent sources collected from published redshift surveys with and without overlap with our sample. The black histograms represent new redshifts from our observations. The line-filled histograms with upper-left orientation represent  targeted sources  from our survey without recognizable spectra. The line-filled histograms with upper-right orientation represent sources without spectroscopic redshift information.  The upper panel indicates  the completeness distribution  of the sample counting sources from our survey and published surveys.}
     \label{fig:hi22}
     \end{figure}
  
  \begin{figure}
   \includegraphics[width=8.8cm]{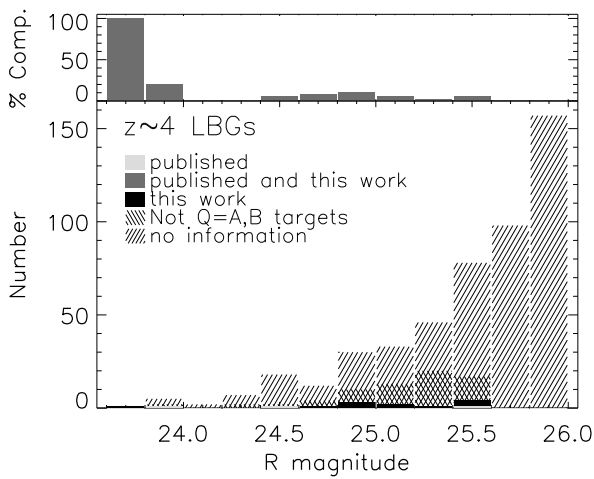}
        \centering
       \caption{Magnitude distribution of $z\sim 4$ optically selected LBG candidates in the SSA22 field. For details about this figure refer to the caption of Figure~\ref{fig:hi22} }
     \label{fig:hi23}
     \end{figure}

\begin{figure}
\includegraphics[width=8.6cm]{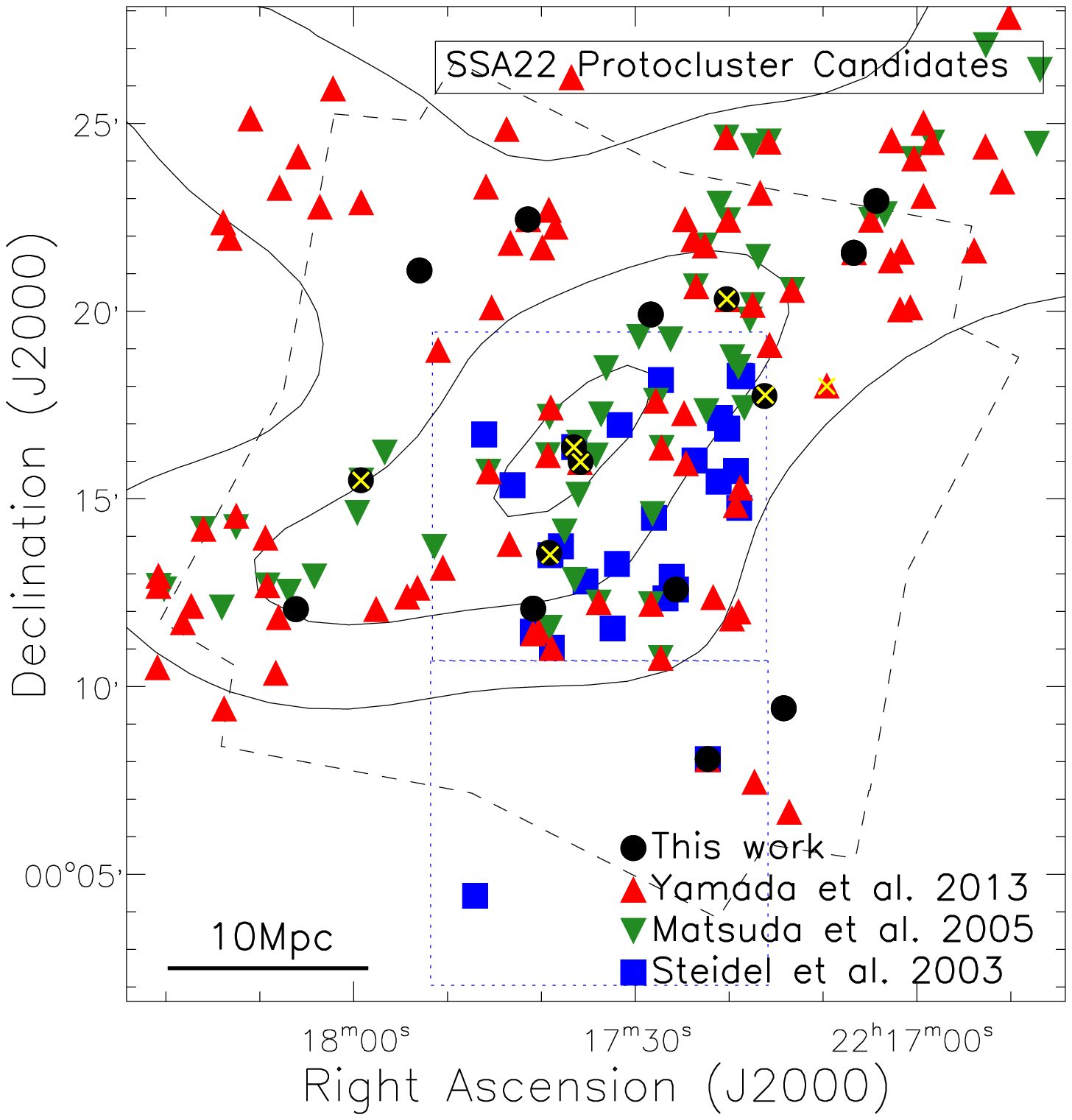}
\caption{Positions of  surveyed sources likely belonging to the SSA22 protocoluster ($3.06 \leq z \leq 3.12$).  Filled circles, triangles, inverted triangles and squares correspond to sources in our survey, the LAE survey of \citet{2012ApJ...751...29Y}, the LAE survey of   \citet{2005ApJ...634L.125M},  and the LBG survey of \citet{2003ApJ...592..728S} respectively. The dashed polygon indicates the area of the \chandra\ SSA22 point source catalog \citep{2009MNRAS.400..299L}. The two dotted squares indicate the SSA22a and SSA22b LBG \citet{2003ApJ...592..728S}  surveys, respectively.  The contours show source isodensity levels of optically selected LAE candidates at $z \sim 3$ from \citet{2004AJ....128.2073H}. The X symbols indicate sources with detected \XR\ emission. The bottom left shows the angular extent of a transverse comoving distance of 10~Mpc at $z=3.091$.}
\label{fig:surv}
\end{figure}
     
In the sample of LBG candidates at $z \sim 3$ and $z \sim 4$ we found eight cases in which the spectrum did not match the typical characteristics of an LBG.  Five out of the eight cases were stars (see Appendix~\ref{S:STAR}); the other three cases include an EllG at $z=0.023$ (10979), an SFG at $z=0.345$ (11281), and an SpG at $z=0.514$ (21493). As a test of possible contamination from low-redshift sources, we tried other templates to fit the LAE or LBG candidates. These alternative templates do not present improved fits over the LBG template.  
For example, the SFG and SpG templates confuses the strong LBG Ly$\alpha$ emission feature with [\ion{O}{ii}]~$\lambda$3727  in the redshift range $0.1\lesssim z \lesssim 1.0$; however, we would expect to observe other features as well (e.g. [\ion{Ne}{iii}]~$\lambda$3864, H$\beta$~$\lambda$4861, [\ion{O}{iii}]~$\lambda$5007) that are not present. Only the spectra of VLT sources 11281 and 21493 exhibit these features.  Indeed, the LBG template does not present any important emission features besides  Ly$\alpha$~$\lambda$1216, though it does have absorption features such as \ion{Si}{ii}~$\lambda$1263, \ion{O}{i}~$\lambda$1304, \ion{C}{ii}~$\lambda$1335 and \ion{Si}{iv}~$\lambda$1397.  With the exception of three spectra (21164, 21186, and 31167; see Figure~\ref{fig:spGA}), there are no clear cases of absorption features in our LBG spectra. In general, our observations are unable to probe the continuum of LBGs. This is expected given that the majority of the optically selected LBGs have $R>24$.  Indeed, S/N in the spectral continua of successfully identified LBGs is low (${\rm \langle S/N\rangle_{cont}} \approx 7$). 

Within the area selected to concentrate this study (see \S  \ref{S:surv}), we check the completeness of the LBG selection by comparing the total number of LBG candidates to the number that have spectroscopic redshifts (see Figures~\ref{fig:hi22} and ~\ref{fig:hi23}). The total number of $z\sim3$ sources with redshifts is 93, with 32 being new redshifts from our survey (see Table~\ref{tab:LBGa}). 
The total number of $z\sim4$ sources with redshifts is 15, with 12 coming from our survey (see Table~\ref{tab:LBGa}).  
Figures~\ref{fig:hi22} and \ref{fig:hi23}, and Table~\ref{tab:LBGa} show that our survey is far from complete, particularly for sources at $z\sim4$. If we include all the spectroscopic information available (including this work), $\approx33\%$ and $\approx14\%$ of the $R<25$ and $R\geq25$ sources have spectroscopic redshifts  for $z\sim 3$ candidates, respectively. Additionally, $\approx10\%$ and $\approx2\%$ of the $R<25$ and $R\geq25$ sources have spectroscopic redshifts  for $z\sim 4$ candidates, respectively. Thus, \RAB\ in our LBG spectroscopic redshift survey was $0.32\pm0.08$ and $0.25\pm0.05$ for the $R<25$ and $R\geq25$  $z\sim 3$ sources, and  $0.27\pm0.11$ and $0.12\pm0.05$ for the $R<25$ and $R>25$ $z\sim 4$ sources (see details in Figures~\ref{fig:hi22} and \ref{fig:hi23}, and Table~\ref{tab:LBGa}), respectively. These numbers suggest that we are only detecting LBGs with strong Ly$\alpha$ emission; in general, our observations are not deep enough to probe the continua of these sources. 

 Similar to this work, \citet{2015A&A...573A..24C} observed high redshift sources with VLT-VIMOS.  Their sample included  
sources with $i' < 25$ and photometric redshifts $z_{\rm phot} =$~2--6.   
 For sources with spectral VVDS flags 3 or 4 (equivalent to $Q$~=~A/B in our survey, see \S\ref{S:resu}), \citet{2015A&A...573A..24C} found 
  \RAB\ fractions of $0.43\pm0.02$ and $0.40\pm0.03$ for $2.7< z_{\rm phot} < 3.5$ and  $3.5< z_{\rm phot} < 4.5$, respectively.     
 These fractions are higher than our values of \RAB\ for $z\sim3$ and $z\sim4$ LBGs with $R<25$. However, these differences are expected given that the \citet{2015A&A...573A..24C} VIMOS masks were observed twice, including one 14 hr exposure using a blue filter ({\sc lrblue}) and one 14 hr exposure using a red filter ({\sc lrred}).  As we discussed earlier in this section, our LBG targets where mainly concentrated in our VLT observations and not in the Keck observations.  Therefore, the \citet{2015A&A...573A..24C} exposures are typically a factor of $\sim$3 times deeper and cover a broader wavelength range than our VLT observations of SSA22 LBGs, which explains the differences in \RAB.




\subsection{New protocluster candidates} \label{S:prot}

We identify new SSA22 protocluster candidates as sources with redshifts $3.06 \leq z \leq 3.12$. This range of redshifts covers a comoving length of $\approx 56$~Mpc. The chosen range more or less agrees with the expected extension of this protocluster, which covers a comoving area of  approximately  60$\times$30~Mpc$^2$ (as shown in \citealt{2004AJ....128.2073H} and Figure~\ref{fig:surv}). Based on this criterion, we found 16 sources in our survey and 134 unique additional sources in published catalogs. We find eight overlapping sources between our observations and the public surveys and hence we report 8 new protocluster candidates (see Figure~\ref{fig:surv}). Most of the protocluster candidates found in our sample (12 out of 16) are from the optically selected LBGs at $z\sim3$ (see \S \ref{S:LBGs}); the rest are two \XR\ selected quasars and two LAE/\XR\ AGNs (see \S \ref{S:XRsr}). The SSA22 protocluster members from the published catalogs come from optically selected LBGs \citep[27 sources from][]{2003ApJ...592..728S} and LAEs \citep[107 sources from][]{2005ApJ...634L.125M, 2012ApJ...751...29Y}. The  protocluster  candidates from  \cite{2003ApJ...592..728S} are primarily from  the SSA22a field, which contains the central region of  the protocluster (see Figure~\ref{fig:surv}). Our criteria to select $z\sim3$ LBGs (see \S\ref{S:LBGs}) are similar to those used by  \cite{2003ApJ...592..728S}; a large fraction ($\approx 48\%$) of \cite{2003ApJ...592..728S} sources are identified by our LBG selection scheme. In contrast, our $z\sim3$ LBG selection criteria is  almost independent of the LAE optical selection criteria of \cite{2004AJ....128.2073H}. For example, for the LAEs  with spectroscopic redshifts, only $\approx 14\%$ of the sources match our selection criteria for $z\sim3$ LBGs.  Additionally, the LAE selection criteria are focused on detecting sources in a narrower range of redshifts (at $z\approx3$) than our $z \sim 3$ LBG selection criteria.  Specifically, the average spectroscopic redshift of the LAEs is $\langle z \rangle = 3.08$ with a standard deviation of 0.02 compared to  an average redshift $\langle z \rangle$ = 3.11 with  standard deviation of 0.21 for the $z \sim 3$ LBGs.  

\subsection{X-ray sources}\label{S:XRsr}

\begin{figure}
   \includegraphics[width=8.95cm]{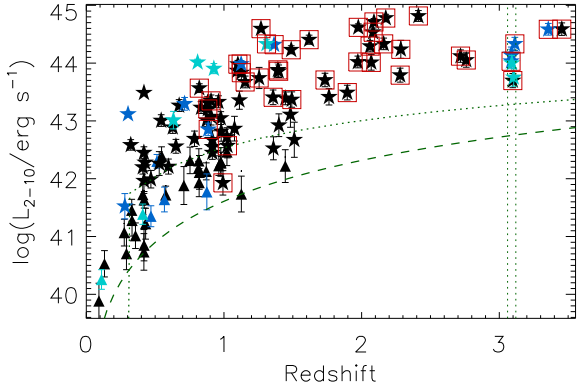}
        \centering
       \caption{X-ray luminosity versus redshift. The 2--10~keV \XR\ luminosities have been determined from the \XR\ fluxes  in \citet{2009MNRAS.400..299L} assuming power-law spectra (see \S \ref{S:XRsr}). 
 The filled stars and triangles indicate sources that have and have not been classified as AGN candidates (according to the selection criteria of \S \ref{S:XRsr}).  
 The black and light-blue sources have redshifts from this work and published data, respectively. Blue symbols indicate sources with redshifts from both our survey and published surveys. Sources that have AGN-type optical spectrum are marked with a square. The dashed curve indicates the sensitivity limit of the \chandra\ SSA22 field survey assuming a power-law spectra with $\Gamma=1.8$ \citep{2009MNRAS.400..299L}. The dotted curve corresponds to the expected quasar sensitivity for a magnitude limit of $R=25.5$.  To obtain this curve, we assumed a power law optical-UV spectrum with $\alpha=-0.5$ \citep[e.g.,][]{2001AJ....122..549V}  and transformations between the optical and X-ray bands following \citet{2004MNRAS.351..169M}.
       The vertical dotted lines indicate the redshift range of the SSA22 protocluster.}     \label{fig:XRay}
     \end{figure}
    
The $\sim 400$~ks \chandra\ Deep Protocluster survey reaches sensitivities of $\sim 3 \times 10^{-16}\, {\rm erg}\, {\rm cm}^{-2}\, {\rm s}^{-1}$ in the observed \HB\ \XR\ band.   
The catalog of \cite{2009MNRAS.400..299L} consists of 297 \XR\ point source detections. For the Keck observations we targeted \XR\ sources by using counterpart positions from the optical/near-infrared observations reported in \citet{2009MNRAS.400..299L}, which are registered to the astrometric frame of the UKIRT Infrared Deep Sky Survey \citep[UKIDSS;][]{2007MNRAS.379.1599L} Deep Extragalactic Survey (DXS) of the SSA22 field. 
In cases where there is not clear $K$-band identifications, we sometimes simply put a slitlet on the \XR\ position while for others we would target a nearby $K$-band source. In the case of the VLT observations we targeted the \XR\ sources when their optical counterparts \citep[from][]{2009MNRAS.400..299L} were detected in the VIMOS pre-image.
We associate a spectrum with an \XR\ source if its angular distance to a \XR\ counterpart is less than $(\sqrt{\Delta \theta^2+2})\arcsec$. The last expression corresponds to the sum of squares of $\Delta \theta$ \citep[the \XR\ positional a error found in Table 2 of][]{2009MNRAS.400..299L} and $\sqrt{2}$ which is an estimation of the expected positional error between optical sources and their \XR\ counterpart.

\begin{table*}
\begin{minipage}{175mm}
\begin{center}
\caption{Spectral properties of extragalactic X-ray sources with redshifts. \label{tab:xray}}
\begin{tabular}{lccccccccccccccccc}
\hline\hline
{\sc xid} &
{\sc id} &
{\sc ref.} &
{\sc RA} &
{\sc DEC} &
$\Delta \theta$  &
$z$  &
$R$  &
$AB_{2500}$  &
${\rm log}~L_{\rm 3.6 \mu m}$ &
${\rm log}~L_{2500}$ &
${\rm log}~L_{2 \rm keV}$ \\
(1) &
(2) &
(3) &
(4) &
(5) &
(6) &
(7) &
(8) &
(9) &
(10) &
(11) &
(12) \\

\hline
    4 &         ... &     5 &  334.24249 &  0.36628 &   1.0 &  0.9266 &   23.4 &  24.49 &  30.80 &  28.11 &  25.94$\pm$0.09\\
    6 &         006 &     1 &  334.24615 &  0.25372 &   1.5 &  1.1261 &   21.9 &  21.79 &  ~...~ &  29.36 &  25.87$\pm$0.13\\
    7 &       10516 &     1 &  334.25137 &  0.33200 &   0.1 &  2.2845 &   21.2 &  20.71 &  31.58 &  30.36 &  26.28$\pm$0.15\\
    8 &         ... &     5 &  334.25208 &  0.35658 &   1.1 &  0.6328 &   23.2 &  23.98 &  29.46 &  27.98 &  25.05$\pm$0.15\\
   10 &         010 &     1 &  334.26251 &  0.25714 &   0.5 &  1.0252 &   24.9 &  25.80 &  29.96 &  27.68 &  24.61$\pm$0.29\\
   11 &       20518 &     1 &  334.27042 &  0.16092 &   0.0 &  2.4122 &   26.9 &  25.75 &  ~...~ &  28.39 &  26.85$\pm$0.09\\
   16 &         016 &     1 &  334.27557 &  0.22722 &   1.2 &  0.9070 &   24.7 &  25.40 &  29.55 &  27.73 &  25.11$\pm$0.14\\
   20 &         ... &     3 &  334.29001 &  0.30003 &   0.9 &  3.1050 &   24.5 &  25.03 &  30.12 &  28.86 &  25.79$\pm$0.21\\
\hline
\end{tabular}
\end{center}
The optical magnitudes used to obtain $AB_{2500}$, $L_{2500}$, \aox, and \lRL\  come from the SDSS survey \citep[e.g.,][]{2008ApJS..175..297A}, the SSA22 photometric survey of \cite{2004AJ....128.2073H} (Subaru magnitudes) and the UKIDSS survey \citep[e.g.,][]{2007MNRAS.379.1599L}. The flux densities at rest-frame 2500\AA\ and 4400\AA\ (from which we obtain $AB_{2500}$, $L_{2500}$, \aox, and \lRL) were obtained by interpolating a power-law fit between two Galactic reddening corrected flux densities obtained from optical magnitudes. If the observed optical flux densities wavelengths do not cover the rest-frame 2500\AA/4400\AA,  the flux densities at rest-frame 2500\AA/4400\AA\ were obtained from the closest wavelength observed flux densities assuming a power-law spectrum $f_\nu\propto \nu^\alpha$ with $\alpha=-0.5$ \citep[e.g.,][]{2001AJ....122..549V}. Table~\ref{tab:xray} is presented in its entirety in the electronic version (also at \url{ftp://cdsarc.u-strasbg.fr/pub/cats/J/MNRAS/450/2615}); an abbreviated version of the table is shown here for guidance as to its form and content. See \S \ref{S:XRsr} for details on the columns and their associated units.  \\

\end{minipage}
\end{table*}

The new spectroscopic classifications for \XR\ sources and their respective \XR\ IDs \citep[from Table 2 of][]{2009MNRAS.400..299L} is presented in Table~\ref{tab:xray} and the details of the columns are given below.

Column (1) gives IDs from Table~2 of \cite{2009MNRAS.400..299L}.

Column (2) gives IDs in case the source belongs to our survey.

Column (3) gives the referred surveys where a source has been spectroscopically observed. 1$\equiv$this work; 2$\equiv$\cite{2003ApJ...592..728S}; 3$\equiv$\cite{2012ApJ...751...29Y}; 4$\equiv$\cite{2004AJ....128..569M}; 5$\equiv$\cite{2005A&A...439..877L}. 

Columns (4) and (5) give the \XR\ point source positions \cite[from][]{2009MNRAS.400..299L} in J2000.0 equatorial coordinates.

Column (6) gives angular distance (in \arcsec) between the \XR\ source and the spectroscopically identified source.

Column (7) gives redshifts, either from our survey or published surveys in case the source is not in our survey.

Column (8) gives the Subaru AB $R$-band magnitudes from the SSA22 photometric survey of \cite{2004AJ....128.2073H}.

Column (9) gives the monochromatic AB magnitude at rest-frame wavelength 2500~\AA. These were computed  by extrapolating the flux densities obtained from the optical magnitudes.  Previous to the extrapolation, the flux densities obtained have been corrected for Galactic reddening.

Column (10) gives the logarithm of the monochromatic luminosity at rest-frame 3.6~$\rm \mu m$ (with units \lumin ${\rm Hz}^{-1}$). These values were computed from the flux densities at rest-frame wavelength 3.6~$\rm \mu m$. The flux densities at rest-frame wavelength 3.6~$\rm \mu m$ have been obtained by interpolating a power-law fit between two Galactic reddening \spitzer\  corrected flux densities (observed at 3.6, 4.5, 5.8 and 8.0~$\rm \mu m$). If the \spitzer\ observed bands do not cover the rest-frame wavelength 3.6~$\rm \mu m$,  the flux densities at rest-frame  3.6~$\rm \mu m$ are obtained from their closest wavelength observed flux densities assuming a power-law spectrum $f_\nu\propto \nu^\alpha$ with $\alpha=-1.0$ \citep[e.g.,][]{2005ApJ...631..163S}.

Column (11) gives the logarithm of the monochromatic luminosity at rest-frame 2500~\AA\ (with units \lumin ${\rm Hz}^{-1}$).

Column (12) gives the logarithm of the monochromatic luminosity at rest-frame 2 keV (with units \lumin ${\rm Hz}^{-1}$) obtained from observed \XR\ fluxes from Table~2 of \cite{2009MNRAS.400..299L}   (see this section for more details).

Column (13) gives the  difference between the logarithm of the monochromatic luminosity at rest-frame 2 keV and the expected value from \cite{2006AJ....131.2826S} (i.e., $\llsoft=0.721\hspace{2pt} \llUV+4.531$).

Column (14) gives the logarithm of the 2--10 keV band luminosity (with units \lumin) obtained from observed \XR\ fluxes of Table~2 of \cite{2009MNRAS.400..299L} (see this section for more details).

Column (15) gives the optical-to-\XR\ power-law slope $\aox = {\rm log}(f_{\rm 2keV}/f_{2500})/ {\rm log}(\nu_{\rm 2keV}/\nu_{2500}$). Where $f_{\rm 2keV}$ and $f_{2500}$ are the flux densities at rest-frame 2~keV and 2500\AA, respectively. 

Column (16) gives the logarithm of the radio-loudness parameter ($R_L = f_{\rm 5 GHz}/f_{4400}$). Where $f_{4400}$ is the flux densities at rest-frame 4400\AA. The flux densities at rest-frame 5~GHz have been obtained using flux densities at 1.4~GHz obtained from the radio images of \cite{2003ApJ...585...57C} assuming a power-law spectrum $f_\nu\propto \nu^\alpha$ with $\alpha=0$.

Column (17) gives 1 if the source is an AGN candidate, 0 otherwise. This is based on our AGN selection criteria described in this section.\\

In the \vlt\ observations, we found reliable redshifts for 35  \XR\ sources. The \keck\ observations exclusively targeted the \XR\ sources, and therefore the majority of this sample  (97 out of 102) correspond to  \XR\ sources. After accounting for the 12 overlapping \XR\ sources between the \keck\ and \vlt\ surveys, in total we found 120 redshifts for \XR\ sources, of which only 15 overlap with redshifts available from published catalogs and 11 are stars. Therefore, we found redshifts for 94 new extragalactic \XR\ sources.  Of the 109 extragalactic \XR\ sources in our survey, 58 are classified as EllGs or SpGs, 9 are SFGs, 3 are LBGs and 39 are AGNs.
Prior to this work there were only 23 \XR\ sources with published spectroscopic redshifts; with this work we increase this to a total of 128 (117 of which are extragalactic sources), an increase by a factor $\approx 5.6$. Note that  \cite{2009MNRAS.400..299L} reports on 46 \XR\ sources with redshifts, though many of these sources are unpublished and from VVDS sources with less secure redshifts than the ones used in this work. In Table~2 of \cite{2009MNRAS.400..299L}  there are 25  redshifts from the unpublished work of Chapman et al. (2009). Of these 25 sources, we obtained redshifts for 13 and in just 4/13 our redshift measurement matches the one provided in Table~2 of  \cite{2009MNRAS.400..299L}. All 13 overlaping sources have $Q$~=~A spectra. Therefore, the Chapman et al. (2009) redshifts do not seem to be robust enough to be used in this work.

Using the redshifts of our survey and those from published catalogs, we obtain X-ray luminosities in the 2--10~keV band $L_{\rm X}$, and monochromatic 2~keV luminosities ($L_{2 \rm keV}$) from the observed \XR\ fluxes found in Table~2 of \cite{2009MNRAS.400..299L}. The X-ray luminosities are estimated  assuming a \PL\ spectrum ($L_\nu=C \nu^{\alpha}$; $\Gamma=1-\alpha$)  with $\Gamma=1.8$  \citep[e.g.,][]{2005MNRAS.364..195P}.  By default the luminosities are obtained by extrapolating  the observed hard-band (\HB) fluxes. In cases where the sources were not detected in the hard-band (\HB), the observed fluxes were obtained in either the soft-band (\SB) or the full-band (\FB).

We classify our \XR\ sample as AGN candidates based on similar criteria to that adopted by  \cite{2012ApJ...752...46L}:
\begin{enumerate}
\item {\it X-ray Luminosity:} for cases in which $L_{X}> 3 \times10^{42}~\lumin$, we can reliably classify the source as an AGN \citep[e.g.,][]{2004AJ....128.2048B}. 

\item {\it X-ray Spectral Shape:} hard \XR\ spectra are indicative of dominant and powerful \XR\ AGN that are significantly obscured. We select such hard AGN candidates based on their hardness ratios, with a cut at an effective $\Gamma < 1$.

\item {\it X-ray-to-Optical Flux Ratio:} \XR-to-optical flux ratios  indicate \XR\ emission which is significantly elevated compared to normal galaxies. We identify AGN candidates based on ${\rm log}(f_X/f_R) > 1.0$  where $f_{\rm X}$ is  either in the HB, FB or SB.\footnote{${\rm log}(f_X/f_R)$ is obtained from ${\rm log}(f_{0.3-3.5}/f_V)={\rm log}~f_{0.3-3.5}+V/2.5+5.37$ \citep[where $f_{0.3-3.5}$ is the 0.3--3.5~keV flux in \flux;][]{1988ApJ...326..680M}. Assuming $V-R=0.22$  which is the expected value for a $\alpha=-0.5$  AGN spectra \citep[e.g.,][]{2000ApJ...532..700A} and correcting $f_{0.3-3.5}$ to the observed flux $f_{\rm X}$ (either at \HB, \FB\ or \SB) assuming $\Gamma=1.8$.}

\item {\it X-ray-to-Radio Luminosity Ratio:} \XR\ emission, as measured by radio emission, is significantly higher than expected from pure star formation; i.e., $L_{\rm 0.5-8 keV} > 5 \times4\times 10^{18}~L_{\rm 1.4GHz}$, where $L_{\rm 1.4GHz}$ is the rest-frame 1.4 GHz monochromatic luminosity in units of ${\rm W}\,{\rm Hz}^{-1}$ and $4\times10^{18}L_{\rm 1.4GHz}$ is the expected \XR\ emission level that originates from starburst galaxies \citep[e.g.,][]{2002AJ....124.2351B, 2003A&A...399...39R,2014MNRAS.437.1698M}.\footnote{From \cite{2002AJ....124.2351B} the following relationship is obtained:  ${\rm log}\,L_{\rm 0.5-8keV}=0.935\,{\rm log}\,L_{\rm 1.4GHz}\,+\,20.141$
 (where $L_{\rm 0.5-8keV}$ is in erg$\,s^{-1}$ and $L_{\rm 1.4GHz}$ is in W$\,$Hz$^{-1}$) and therefore in the range of radio luminosities where this relationship is valid (i.e. $19\!\lesssim\!{\rm log}\,L_{\rm 1.4GHz}\!\lesssim 24$) $L_{\rm X}\sim 4\!\times\!10^{18} L_{\rm 1.4 GHz}$; ($L_{\rm X}\approx0.68\!\times\!L_{\rm 0.5-8keV}$ for a power-law spectrum with $\Gamma=1.8$).} The radio fluxes are obtained from the SSA22 VLA radio images of \cite{2003ApJ...585...57C} with an assumed spectral index of $\alpha_r=0$ \citep{1997ApJ...475..479W}; $L_{\rm 0.5-8 keV} $ is obtained from $L_{\rm X}$ assuming a power-law spectra with $\Gamma=1.8$.

\item {\it Optical spectroscopy:} Based on our source classification for sources is in our spectroscopic survey. For the few cases that  sources are not in our spectroscopic sample, we search for evidence of AGN features in the literature if applicable.
\end{enumerate}
Our criteria indicate 84 (out of 117) of the \XR\ extragalactic sources  are AGNs.
In our optical spectroscopy, only 39 out of 78 of the AGNs candidates were spectroscopically classified as AGNs. 
The lack of AGN features in the optical spectra in half of the \XR\ AGNs is expected from AGN obscuration \citep[e.g.,][]{2001AJ....122.2156A, 2005AJ....129..578B}. 
In Figure~\ref{fig:XRay} we show the \XR\ luminosity  versus redshift for the \XR\ sources. Note that the sensitivity limit of the SSA22 \chandra\ survey implies that every \XR\ source above $z \gtrsim 1.5$ is an AGN. 

We find 7 sources that potentially belong to the SSA22 protocluster ($3.06 < z < 3.12$). 
These sources have \XR\ luminosities $L_{2-10} \gtrsim 5\times10^{43}~\lumin$ implying they are  quasars\footnote{Quasars are defined here as AGNs with with bolometric luminosities $L_{\rm bol} \gtrsim 10^{44}~\lumin$.} with $L_{\rm bol} \gtrsim 10^{45}~\lumin$ \citep[e.g.,][]{2004MNRAS.351..169M}. The SSA22 protocluster candidates with \XR\ detections are marked with X symbols in Figure~\ref{fig:surv}.  
In the work of  \cite{2009MNRAS.400..299L} there are 9 reported \XR\ sources at $3.06 \lesssim z \lesssim 3.12$; however, 3 of the 9 sources came from the unpublished work of Chapman et al. (2009).  As mentioned earlier, we do not adopt  Chapman et al. (2009) redshifts in this work.
Of the seven \XR\ protocluster candidates, five are from our spectroscopy (one new source found in our survey),  and three show QSO signatures in their optical spectra. Our sample of X-ray protocluster candidates might not be complete. The missing sources could be missed targets or cases with fainter fluxes than the sensitivity limits of our survey. For example the X-ray source 120 in the Lehmer et al. catalog, has a photometric redshift within the range $3.06 \lesssim z \lesssim 3.12$ \citep{2010ApJ...724.1270T}. This source was targeted by our survey, but the resulting spectrum did not show any recognizable feature. Additionally, observations of this source at sub-millimeter wavelengths complemented by a hard photon index ($\Gamma =-0.3 \pm 0.6$) strongly indicate an obscured protocluster quasar \citep{2010ApJ...724.1270T}.
From the photometric redshifts of \cite{2013ApJ...778..170K}, we also expect to find  additional X-ray protocluster members. In the mentioned work, it is found that there are 19 X-ray sources within $2.6 < z < 3.6$. In contrast, we found 11 sources in this redshift range, and therefore, more X-ray protocluster members with reasonable optical magnitudes for spectroscopy may exist.
\begin{figure}
   \includegraphics[width=8.8cm]{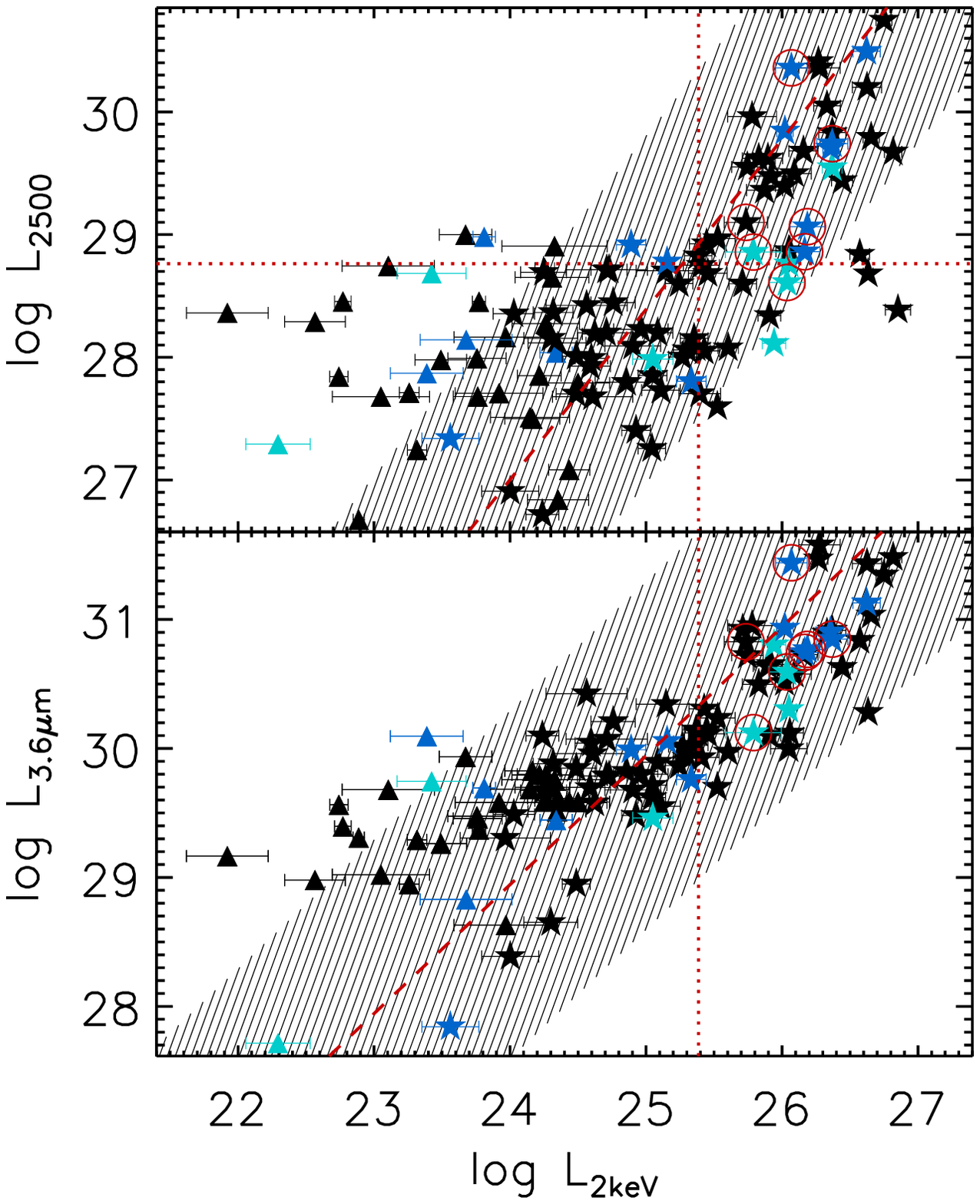}
        \centering
       \caption{Infrared 3.6~$\rm \mu m$ (lower panel) and ultraviolet 2500~\AA\  (upper panel) luminosity versus 2~keV luminosity for \XR\ sources in the \chandra\ Deep Protocluster survey. Monochromatic luminosities used here have units of $\lumin {\rm Hz}^{-1}$. Circles indicate sources that potentially belong to the SSA22 protocluster ($3.06 \leq z \leq 3.12$). The dashed lines in the lower and upper panel are the linear relations estimated from \citet{2006AJ....131.2826S} and \citet{2006ApJS..166..470R} and the shadow regions are their respective 2-$\sigma$ confidence limits. The dotted vertical lines in the lower and upper panels are the sensitivity limits of the  SSA22 \chandra\ survey at $z=3.091$ assuming a power-law spectrum with $\Gamma=1.8$. The dotted horizontal line in the upper panel is the $R=25.5$ sensitivity limit at $z=3.091$ assuming a power-law optical-UV spectrum with $\alpha=-0.5$ \citep[e.g.,][]{2001AJ....122..549V}.
        Symbols and colors of the sources are described in the caption of Figure~\ref{fig:XRay}.}     \label{fig:IRXR}
     \end{figure}
     
To check additional properties of the \XR\ sample, we plot  the 2~keV luminosity versus the 3.6~$\rm \mu m$ luminosity and  the  2~keV luminosity versus the 2500~\AA\ luminosity in Figure~\ref{fig:IRXR}. With the exception of some outliers, our AGN candidates satisfy the $L_{2500}-L_{\rm 2keV}$ relation of \cite{2006AJ....131.2826S}. The  \lIR\ to \lsoft\  ratio of \XR\ AGN candidates seem to match the expected values when compared to the  \cite{2006ApJS..166..470R} SEDs (see lower panel Figure~\ref{fig:IRXR}).

In this work we obtained the radio loudness parameter by extrapolating optical and radio flux densities. The radio flux densities were sensitive enough\footnote{The limiting 1.4~GHz sensitivity is 60~$\mu$Jy \citep[5$\sigma$;][]{2003ApJ...585...57C}.} to properly discriminate between radio-loud (those with $\lRL> 1.0$) and radio-quiet (those with $\lRL \leq 1.0$) AGNs for a fraction of our sample. In our \XR\ extragalactic sample 12 sources (nine AGNs) have radio detections with four of them corresponding to radio-loud AGN with moderate radio loudness parameters ($1.2 \lesssim \lRL \lesssim 1.6$). Additionally, at least 23 (four protocluster candidates) out of the 75 AGNs that do not have radio detection cannot be excluded  (with $\lRL$ upper limits $\geq$1.5) from being radio-loud (see Table~\ref{tab:xray}).


 
\subsection{Peculiar spectra in the SSA22 Field} \label{S:PeSp}

In this section we describe  three \XR\ sources with peculiar spectrum. 

\begin{figure}
   \includegraphics[width=8.8cm]{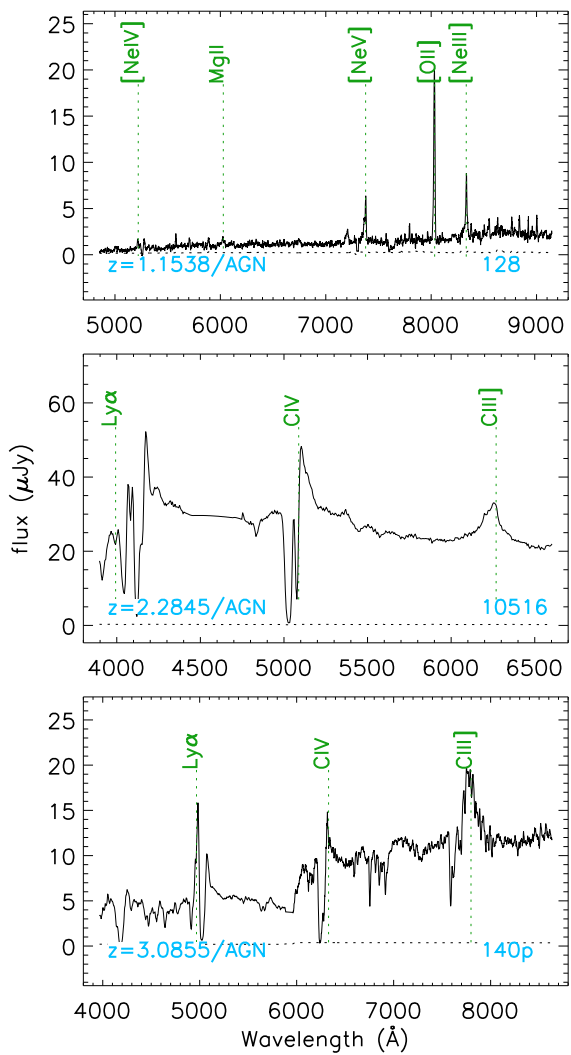}
        \centering
       \caption{Three peculiar spectra in the SSA22 field. In each figure key emission lines of the AGN templates have been marked for reference.}     \label{fig:PECS}
     \end{figure}

\subsubsection{X~128:  A Ne-Rich Type-2 AGN at $z=1.154$}

X~128 is a moderate-redshift type-2 AGN showing an interesting set
of emission lines at $z = 1.154$, with line ratios atypical of
composite obscured AGN spectra (see Figure~\ref{fig:PECS}).  Specifically, several
ionization states of neon are particularly strong, including
[\ion{Ne}{iv}]~$\lambda$2424, [\ion{Ne}{iii}]~$\lambda$3343,
[\ion{Ne}{v}]~$\lambda$3346, [\ion{Ne}{v}]~$\lambda$3426,
[\ion{Ne}{iii}]~$\lambda$3869, and [\ion{Ne}{iii}]~$\lambda$3968.  In
radio galaxy spectra, typically [\ion{Ne}{iv}]~$\lambda$2424 is
observed with comparable strength to \ion{C}{ii}]~$\lambda$2326
\citep[e.g.,][]{1993ARA&A..31..639M, 1999AJ....117.1122S}, whereas here the [\ion{Ne}{iv}]
line appears at least $\gtrsim 10\times$ the \ion{C}{ii}] line strength.   The observed lines are not unprecedented, however; for
instance, these ionization states are all evident in {\it Hubble
Space Telescope} Imaging Spectrograph (STIS) observations of the
``hot spot'' in the inner narrow-line region of the well-studied
Seyfert~2 galaxy NGC~1068 \citep{1998ApJ...508..232K, 2000ApJ...532..256K}.  In that
source, high S/N spectra, including several dozen emission lines
from the UV through the optical, are modeled as coming from multiple
photoionized gas components.

\subsubsection{BAL quasars}

In this work we found two BAL quasar spectra: VIMOS 10516 at $z=2.285$ and LRIS 140p at $z=3.086$ (see Figure~\ref{fig:PECS}). Both sources present Ly$\alpha$~$\lambda$1216, \ion{C}{iv}~$\lambda$1549 and \ion{C}{iii}]~$\lambda$1909  emission lines.  These BAL quasars show \ion{C}{iv} blue-shifted absorption with projected  line-of-sight Doppler broadenings $\sim -5500$~\kms\ and $\sim -5000$~\kms\ for 10516 and 140p, respectively.
These sources as expected present moderate-to-weak \XR\ emission with respect to normal quasars \citep[e.g.,][]{2006ApJ...644..709G, 2009ApJ...692..758G, 2012ApJ...759...42S} with $\Delta \llsoft=-0.15$ and $-0.35$ for  10516 and 140p, respectively; where $\Delta \llsoft$ is the difference between \llsoft\ and its expected value from the $L_{2500}-L_{\rm 2keV}$ relation of \cite{2006AJ....131.2826S}.

\section{SUMMARY AND FUTURE WORK} \label{S:conc}

In this paper we present a survey of extragalactic sources in the SSA22 field obtained using multi-object spectrographs on the VLT and Keck telescopes.
We quantify redshifts for new LBGs in the redshift range of the SSA22 protocluster and \XR\ sources in the \chandra\ Deep Protocluster Survey with the goals of finding new protocluster members  and complementing previous studies performed in this field. The main results can be summarized as follows:

\begin{enumerate}
\item{By template matching high quality \vimos, \deimos, and \lris\ spectra we  have successfully identified redshifts for 247 extragalactic sources and 120 Galactic stars. Discounting matches with sources that have redshift identifications from published catalogs, we found redshifts for a total of 206 new extragalactic sources.}

\item{We have substantially increased  the number of sources with known redshifts in the redshift range between $1.0 \leq z \leq 2.0$ and $z \geq 3.4$, by factors of $3.5$ and $13$, respectively.}

\item{By targeting LAE and LBG candidates, we have successfully identified redshifts for 36 new sources at  $z>2$ with ${\rm Ly}\alpha$ spectral features. All but one source has strong ${\rm Ly}\alpha$ emission line; the exception (VLT  21164) has ${\rm Ly}\alpha$ absorption. }

\item{We have identified 8 new SSA22 protocluster candidates ($3.06 \leq z \leq 3.12$),  one of of which is an \XR\ detected quasar.}

\item{We have increased the number of \XR\ sources with reliable redshifts from 23 to 128, leaving the \cite{2009MNRAS.400..299L} catalog now $\approx 43$\% complete.}

\item{Using a variety of multiwavelength criteria, we identified 84 out of the 128 \XR\ sources in the SSA22 field as AGNs, with 7 AGNs protocluster candidates (one new from this work; $3.06 \leq z \leq 3.12$). These candidates correspond to quasars with bolometric luminosities $L_{\rm bol} \gtrsim  10^{45}~\lumin$.}

\end{enumerate}

The spectroscopic survey performed in this work is far from complete. There is still an important fraction of \XR\ sources without any spectral identification ($\approx 57\%$). Additionally, there are $184$ optically identified $R<25$ LBGs candidates without reliable redshifts. The level of completeness of LBGs at $z\sim3$ and $z\sim4$ with $R<25$ is $\approx33\%$ and $\approx10\%$, respectively  (see \S\ref{S:LBGs}).   
Since most optically selected LBGs have $R > 24$, in general, our observations are neither probing  the continuum nor the absorption spectra of LBGs. More and deeper spectral surveys are needed both to find a more complete sample of the SSA22 protocluster and to fully exploit the $\sim 400$~ks \chandra\ observation available in this field.

\appendix
\renewcommand\thefigure{A.\arabic{figure}}
\renewcommand\thetable{A.\arabic{table}}
\setcounter{figure}{0}
\setcounter{table}{0}

\begin{figure*}
\includegraphics{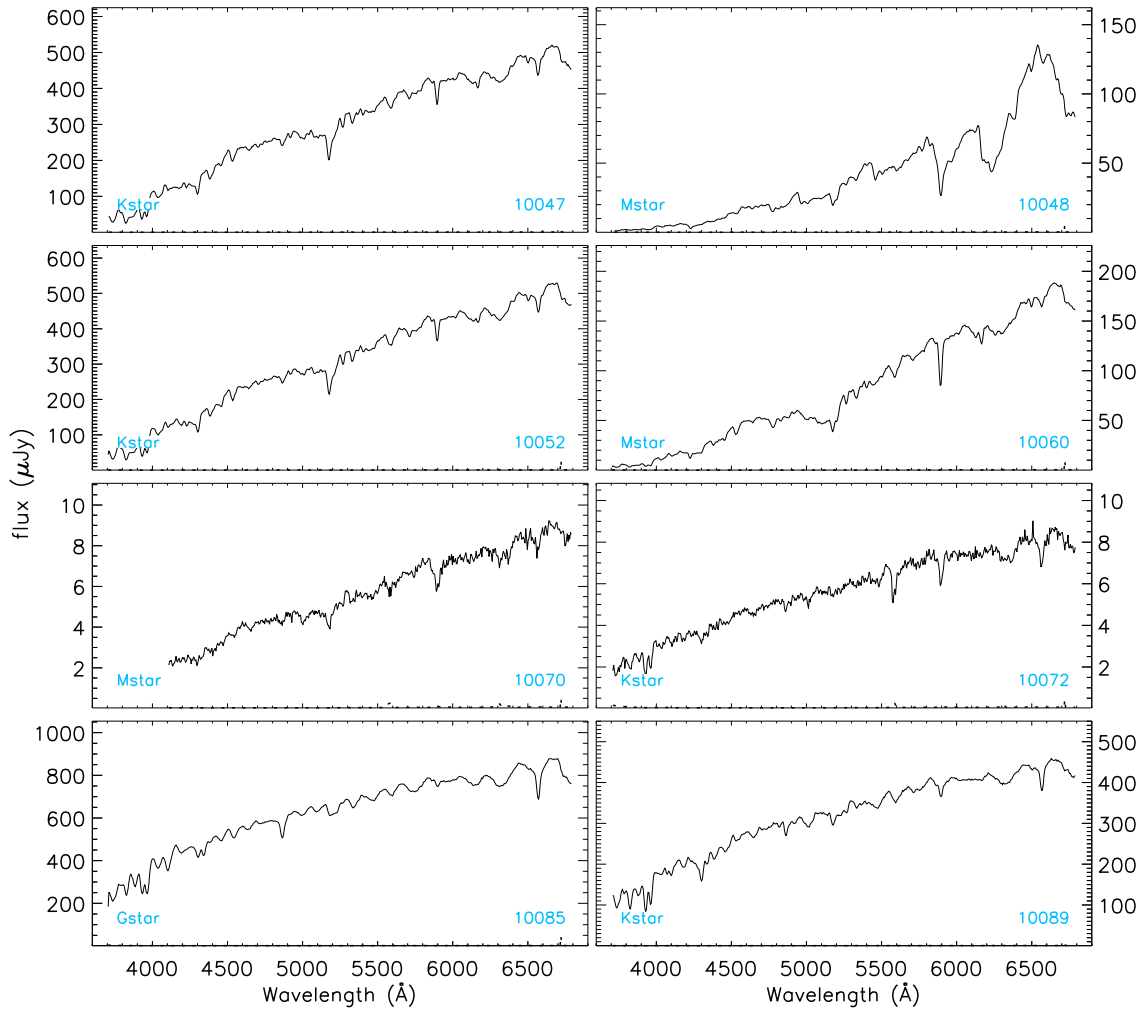}
\caption{\vimos\ stellar spectra.  For display purposes only the first eight \vimos\  spectra are shown. Please refer to the electronic version to see all the spectra.}
\label{fig:sSTA}
\end{figure*}

\begin{figure*}
\includegraphics{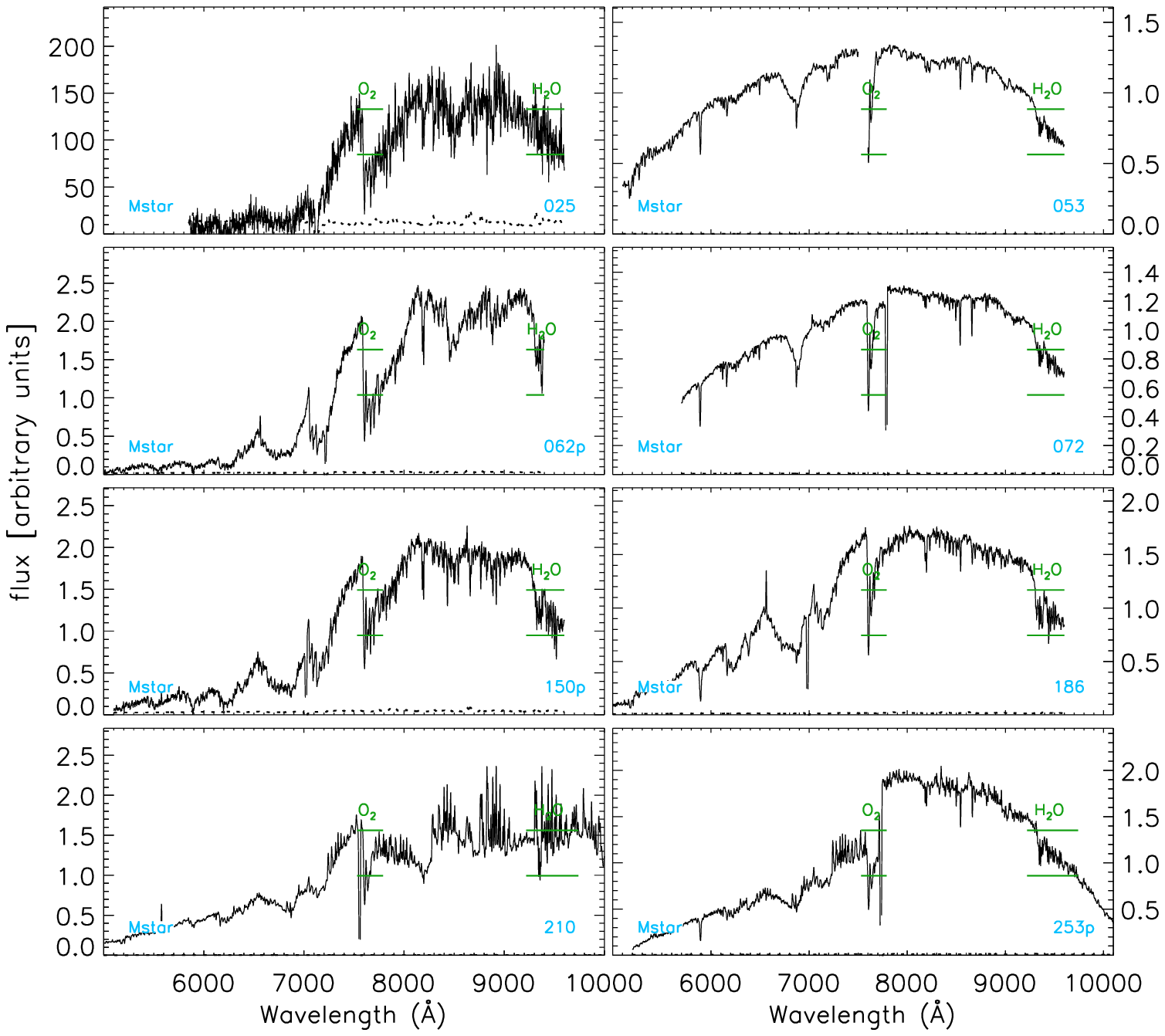}
\caption{\deimos\ and \lris\ stellar spectra.}
\label{fig:sST2}
\end{figure*}

\begin{table*}
\begin{minipage}{175mm}
\begin{center}
\caption{Spectroscopically surveyed Galactic stars. \label{tab:spST}}
\begin{tabular}{ccccccccccc}
\hline\hline
{\sc object id} &
{\sc RA$^{\rm a}$} &
{\sc DEC$^{\rm a}$} &
{\sc sdss $g^{\rm b}$} &
{\sc subaru $R^{\rm b}$} &
{\sc ukidss $K^{\rm b}$} &
{\sc s. type$^{\rm c}$}  &
{\sc masks$^{\rm d}$}  &
{\sc quality} &
{\sc class$^{\rm e}$} \\
\hline
\\
\multicolumn{10}{c}{\small  VLT observations} \\
 \\
\hline
   10047 & 334.49319 &  0.14679 &  18.3 &  18.3 &  17.4 &   Kstar &     1 &  A &           bri \\
   10048 & 334.52917 &  0.14442 &  20.9 &  19.4 &  17.2 &   Mstar &   1,4 &  A &           bri \\
   10052 & 334.53781 &  0.14977 &  18.2 &  18.2 &  17.3 &   Kstar &   1,4 &  A &           bri \\
   10060 & 334.52075 &  0.15687 &  20.0 &  18.9 &  17.8 &   Mstar &   1,4 &  A &           bri \\
   10070 & 334.27811 &  0.16194 &  22.5 &  22.0 &  21.7 &   Mstar &   1,4 &  B &           bri \\
   10072 & 334.35138 &  0.16143 &  22.3 &  22.0 &  22.1 &   Kstar &   1,4 &  B &           bri \\
   10085 & 334.43250 &  0.17121 &  17.4 &  18.4 &  17.6 &   Gstar &   1,3 &  A &           bri \\
   10089 & 334.33994 &  0.17206 &  18.1 &  18.6 &  18.0 &   Kstar &     1 &  A &           bri \\
\hline
\\
\multicolumn{10}{c}{\small  Keck observations} \\
 \\
 \hline
 
     025 & 334.29446 &  0.18511 &   ... &  24.8 &  21.9 &   Mstar &     3 &  A &            XR \\
     053 & 334.32986 &  0.24122 &  17.6 &  18.4 &  16.1 &   Mstar &     3 &  A &        XR/bri \\
    062p & 334.33701 &  0.17625 &  24.0 &  22.8 &  19.1 &   Mstar &     2 &  A &           ... \\
     072 & 334.34821 &  0.24539 &  21.2 &  20.0 &  18.9 &   Mstar &     3 &  A &        XR/bri \\
    150p & 334.41080 &  0.26592 &  24.6 &  23.0 &  20.4 &   Mstar &     1 &  A &           ... \\
     186 & 334.43619 &  0.29689 &  22.8 &  21.2 &  19.4 &   Mstar &     2 &  A &        XR/bri \\
     210 & 334.44656 &  0.28000 &  17.4 &  18.5 &  14.2 &   Mstar &     1 &  B &        XR/bri \\
    253p & 334.47568 &  0.37325 &  21.0 &  19.6 &  18.1 &   Mstar &     1 &  A &           bri \\
     \hline
\end{tabular}
\end{center}

$^{\rm a}$Optical positions in J2000.0 equatorial coordinates. \\
$^{\rm b}$The optical magnitudes presented in this Table come from the SDSS survey \citep[e.g.,][]{2008ApJS..175..297A}, SSA22 photometric survey of \cite{2004AJ....128.2073H} (Subaru magnitudes) and the UKIDSS survey \citep[e.g.,][]{2007MNRAS.379.1599L}. The UKIDSS magnitudes have been transformed from Vega to AB magnitudes using K(AB)=K(Vega)+1.9 \citep[from][]{2006MNRAS.367..454H}.  \\
$^{\rm c}$ Based on the cross-correlation star template that has minimum $\chi^2$. \\
$^{\rm d}$ Masks where source presents A or B quality spectra.\\
$^{\rm e}$ Target selection criteria classes. bri $\equiv$ bright source ($R<22.5$); XR $\equiv$ X-ray source from the \cite{2009MNRAS.400..299L} catalog; $z$$\sim$3 $\equiv$ $z\sim3$ LBG; $z$$\sim$4 $\equiv$ $z\sim4$ LBG; Ste03 $\equiv$ LBG from \cite{2003ApJ...592..728S}; LAE $\equiv$ LAE from \cite{2004AJ....128.2073H}.  \\
Table~\ref{tab:spST} is presented in its entirety in the electronic version (also at \url{ftp://cdsarc.u-strasbg.fr/pub/cats/J/MNRAS/450/2615}); an abbreviated version of the table is shown here for guidance as to its form and content. 
\end{minipage}
\end{table*}

\section*{Acknowledgements}

We would like to thank Ezequiel Treister and the anonymous
referee for helpful discussions regarding the interpretations of our results. 
We would also like to thank Scott Chapman for  providing radio images of the SSA22 field.
CS acknowledges support from CONICYT-Chile (FONDECYT 3120198, Becas Chile 74140006, and the Anillo ACT1101).
FEB acknowledges support from CONICYT-Chile 
(Basal-CATA PFB-06/2007, FONDECYT 1141218, Gemini-CONICYT 32120003, "EMBIGGEN" Anillo ACT1101),
and Project IC120009 "Millennium Institute of Astrophysics (MAS)" funded
by the Iniciativa Cient\'{\i}fica Milenio del Ministerio de
Econom\'{\i}a, Fomento y Turismo. The work of DS was carried out at Jet Propulsion
Laboratory, California Institute of Technology, under a contract
with NASA. JEG thanks the Royal Society.

Based on observations made with ESO Telescopes at the Paranal Observatory under programme IDs 085.A-0616 and 089.A-0405. Some of the data presented herein were obtained at the W.M. Keck Observatory, which is operated as a scientific partnership among the California Institute of Technology, the University of California and the National Aeronautics and Space Administration. The Observatory was made possible by the generous financial support of the W.M. Keck Foundation.ÊThe authors wish to recognize and acknowledge the very significant cultural role and reverence that the summit of Mauna Kea has always had within the indigenous Hawaiian community.Ê We are most fortunate to have the opportunity to conduct observations from this mountain.

\bibliographystyle{apj}

\providecommand{\noopsort}[1]{}

\section{Spectra of Stars in the SSA22 Field \label{S:STAR}}

As for extragalactic sources, Galactic stars were recognized by cross correlating template spectra with the data using the {\sc specpro} tool. The templates used \citep[from][]{1998PASP..110..863P} allow us to get a reliable stellar classification based on main-sequence classes (A, F, G, K, and M stars). To classify stars we choose the template with minimum $\chi^2$. We found 113 stars in the VLT observations (see Table~\ref{tab:spST} and Figure~\ref{fig:sSTA}) and 8 stars in the Keck observations  (see Table~\ref{tab:spST} and Figure~\ref{fig:sST2}). There is one star which was observed by both VLT and Keck, corresponding to 20644 (VLT) and 186 (Keck), such that we have surveyed a total of 120 Galactic stars.
The majority are K and M stars (110/120) but we also find a few G and F stars (10/120).

\label{lastpage}

\end{document}